\documentclass[twocolumn, twocolappendix]{aastex701}

\submitjournal{ApJ}

\shorttitle{ISM-Driven Radial Heating and Migration}
\shortauthors{Modak et al.}
\graphicspath{{./}{figures/}}

\usepackage{graphicx}
\usepackage{bm}
\usepackage{hyperref}
\usepackage{float}
\usepackage{amsmath, amssymb}
\usepackage{xcolor}
\usepackage{enumitem}
\usepackage{comment}

\usepackage[normalem]{ulem}

\newcommand{\Msun}{M_\odot}
\newcommand{\md}{\mathrm{d}}
\newcommand{\bx}{\bm{x}}
\newcommand{\bvee}{\bm{v}}
\newcommand{\bk}{\bm{k}}

\newcommand{\bJ}{\bm{J}}
\newcommand{\btheta}{\bm{\theta}}
\newcommand{\nn}{\nonumber}

\newcommand{\ztilde}{\tilde{z}}
\newcommand{\ktilde}{\tilde{k}}

\newcommand{\sech}{\mathrm{sech}}
\newcommand{\p}{\partial}
\newcommand{\me}{\mathrm{e}}

\newcommand\CHrem{\bgroup\markoverwith{\textcolor{magenta}{\rule[0.5ex]{2pt}{0.9pt}}}\ULon}

\begin{document}

\title{Interstellar Medium-Driven Orbital Transport -- I. Radial Heating and Migration}

\author[0000-0002-8532-827X]{Shaunak Modak}
\affiliation{Department of Astrophysical Sciences, 4 Ivy Lane, Princeton University, Princeton, NJ 08544, USA}
\email{shaunakmodak@princeton.edu}

\author[0000-0002-5861-5687]{Chris Hamilton}
\affiliation{Department of Astrophysical Sciences, 4 Ivy Lane, Princeton University, Princeton, NJ 08544, USA}
\affiliation{School of Natural Sciences, Institute for Advanced Study, Einstein Drive, Princeton NJ 08540}
\email{chamilton@ias.edu}

\author[0000-0002-0509-9113]{Eve C. Ostriker}
\affiliation{Department of Astrophysical Sciences, 4 Ivy Lane, Princeton University, Princeton, NJ 08544, USA}
\email{eostrike@princeton.edu}

\author[0000-0002-0278-7180]{Scott Tremaine}
\affiliation{School of Natural Sciences, Institute for Advanced Study, Einstein Drive, Princeton NJ 08540}
\affiliation{Canadian Institute for Theoretical Astrophysics, University of Toronto, 60 St. George Street, Toronto, ON M5S 3H8, Canada}
\email{tremaine@ias.edu}

\begin{abstract} 
Interstellar medium (ISM) structures gravitationally perturb stellar orbits in galactic disks, driving orbital heating and migration. However, studies of these transport processes tend to model the ISM very crudely, e.g., as a collection of compact, spherical ``clouds'' moving in the disk plane. Here, we revisit this problem with more realistic models of ISM density fluctuations drawn from the TIGRESS-NCR magnetohydrodynamic simulations, which follow the physics governing the ISM in Milky-Way-like conditions at high resolution. By integrating test-particle trajectories through time-dependent TIGRESS-NCR structures, we uncover transport behavior that contrasts sharply with conventional theoretical expectations. Notably, radial heating scales as $\sigma_R \propto t^{1/2}$ for initially cold orbits at early times, and $\sigma_R \propto t^{1/5}$ for warmer orbits at late times, contrary to the classic $\sigma_R \propto t^{1/3}$ prediction. The ISM drives substantial radial migration, accounting for $\gtrsim 30\%$ of that observed in the solar neighborhood (even without stellar spiral structure), and leads to a very low heating-to-migration ratio of $\mathrm{rms}\,\delta J_R\,/\,\mathrm{rms}\,\delta J_\varphi \approx 0.055$, where $J_R$ and $J_\varphi$ are the radial and azimuthal actions respectively. Vertical motion suppresses the amplitude of radial transport, but does not change the basic scalings. All our simulation results can be explained using quasilinear diffusion theory, accounting for the fact that the dominant ISM fluctuations have wavelengths of $\lambda_* \sim 600\,$pc and correlation timescales of $\tau_* \sim 70\,$Myr. We provide simple fitting formulae for the corresponding diffusion coefficients. In Paper II, we study the ISM's role in vertical disk heating.
\end{abstract}


\section{Introduction}
\label{sec:introduction}

The interstellar medium (ISM) is an important driver of secular evolution in the stellar disks of galaxies. In typical present-day galaxies, the ISM constitutes roughly $10\%$ of the baryonic disk mass \citep{SaintongeCatinella2022}, and makes up an even larger fraction of disk mass in high-redshift galaxies \citep{TacconiGenzelSternberg2020}. The ISM exhibits substantial spatial structure (see, e.g.,  reviews by \citealt{HeyerDame2015, Hacar2023, McClureGriffiths2023, SchinnererLeroy2024}), and the gravitational potential fluctuations associated with this structure continually perturb stellar orbits. The cumulative effect of such perturbations is \textit{orbital transport}, i.e., the gradual redistribution of stars throughout orbital phase space \citep{BinneyLacey1988, Sellwood2014}. In galactic disks, this transport is typically decomposed into three components: radial heating (an increase in a stellar population's radial velocity dispersion), vertical heating (an increase in a stellar populations' vertical velocity dispersion, and the stellar disk scale height), and radial migration (changes in a star's orbital angular momentum or guiding-center radius).

Secular evolution arising from orbital transport is probed observationally via variations in orbital element distributions, with chemical composition tagging coevally and cospatially formed populations (see, e.g., \citealt{BlandHawthorn2010, Hogg2016}). In general, stellar age and metallicity distributions vary with both radial and vertical location and kinematics in galactic disks, and these may be characterized for both the Milky Way and external galaxies.

For the Milky Way, modern surveys such as \textit{Gaia} \citep{GaiaDR3}, DESI \citep{DESI}, and the Sloan Digital Sky Survey V (SDSS-V)'s Milky Way Mapper \citep{SDSSV} are providing increasingly precise measurements of the phase-space structure of the disk through 6D kinematics and chemical abundance measurements of individual stars (see \citealt{hunt2025milky} for a recent review), and these data in turn enable precise constraints on the Galaxy's chemo-dynamical history (e.g., \citealt{ChibaSchonrich2021, Asano2025, Chiba2026}). In external systems, integral field spectroscopic surveys like MaNGA \citep{MaNGA} probe 3D projected phase space (i.e., on-sky positions and line-of-sight velocity distributions), enabling similar chemo-dynamical studies using stellar population synthesis and orbit superposition models. Radial heating, vertical heating, and radial migration form a major part of this observational story (see, e.g., \citealt{Mackereth2019, Frankel2020, Zhang2025data} for Milky-Way studies and \citealt{Shetty2020} for an analysis of several Milky-Way-like MaNGA galaxies).

A comprehensive understanding of the physics governing these transport processes remains outstanding. However, as we will show in this paper, we expect that gravitational perturbations from ISM density fluctuations play a significant role. Moreover, observations from facilities including the James Webb Space Telescope (JWST) are uncovering gas-rich galactic disks at high redshifts (e.g., \citealt{TsukuiIguchi2021, Rowland2024, Wang2025}), where the ISM's role in shaping disk evolution was surely even more pronounced. To interpret any of these data accurately and infer the Galaxy's dynamical history, including the possible role of dark matter substructure, it is essential that we possess sophisticated models of ISM-driven orbital transport. Unfortunately, while the potential importance of the ISM has long been appreciated, realistic models are currently lacking, and many stellar-dynamical studies still rely on highly idealized models of the ISM, or even omit it altogether.

The possibility that ISM structures could drive significant transport was first recognized in a pair of seminal papers by \cite{SpitzerSchwarzschild1951, SpitzerSchwarzschild1953}. In particular, \cite{Spitzer1948} summarized contemporary evidence (from dust extinction and sodium absorption lines) for a population of large and dense ``interstellar clouds,'' with \cite{SpitzerSchwarzschild1951, SpitzerSchwarzschild1953} arguing that if the largest clouds are sufficiently massive, gravitational encounters with them could account for the observed stellar age-velocity dispersion relation (AVR) in the solar neighborhood. Only much later were the masses and spatial distribution of giant molecular clouds (GMCs) characterized based on CO observations (e.g., \citealt{ScovilleSolomon1975, Solomon1987}). Subsequent to the work of Spitzer and Schwarzschild, a number of analytic studies have advanced our understanding of transport by such ``ISM clouds.'' \cite{Julian1967} applied conservation of the Jacobi integral of motion in the frame corotating with an individual perturber to note that radial heating and migration are directly linked, with the cloud simply mediating the exchange of energy associated with the star's azimuthal (``circular'') and radial (``disordered'') motion. These analyses were generalized to incorporate finite cloud lifetimes by \cite{Fujimoto1980}, and from two dimensions to three by \cite{Lacey1984, Ida1993}. \cite{Carlberg1987} emphasized that the ISM likely plays an important role in disk thickening. Works by \cite{BinneyLacey1988, JenkinsBinney1990, Jenkins1992} concluded that the ISM is likely responsible for much of the vertical heating in disks, but radial heating and migration should be attributed to spiral wave perturbations instead.

Subsequent $N$-body studies modeling the ISM as, e.g., a fixed collection of Plummer-sphere (or similar) perturbers by \cite{KokuboIda1992, ShiidsukaIda1999, HanninenFlynn2002}, as well as more sophisticated simulations by \cite{AumerBinneySchonrich2016a, AumerBinneySchonrich2016b} (incorporating finite cloud lifetimes and power-law mass distributions) are largely in agreement with the aforementioned analytic expectations. These works have also motivated several cloud-based semi-analytic models for ISM scattering processes (e.g., \citealt{Wu2022, StruckElmegreenDOnghia2026}) which yield the same transport outcomes and may be compared with local stellar kinematics without incurring as much computational cost.

However, a key shortcoming of all these works is their reliance on essentially the same ``collection of GMCs'' model for ISM structure that was posited in the 1950s. While this model is convenient because it often leads to analytically tractable results, it is fundamentally unrealistic. First of all, we know that the ISM does not consist of isolated spherical clouds but rather a hierarchy of extended structures at multiple scales that are often filamentary in morphology (e.g., \citealt{Hacar2023} and references therein). In addition, it seems likely that spherical cloud models lead to distinct transport behavior from fluctuations self-consistently generated by solving the time-dependent (magneto-)hydrodynamic problem, as attested by recent global simulations that include both a stellar and gaseous component (e.g., \citealt{Fujimoto2023, Arunima2025, Zhang2025}). While these simulations represent a significant step forward compared to previous works, they are themselves often limited for computational reasons. Typically, such global simulations ignore magnetic fields, and do not have sufficient spatial and temporal resolution to model the detailed stellar feedback, chemistry, and cooling required for a realistic multiphase ISM. Without the energy input from \ion{H}{2} regions and supernovae that drives dynamics in the ISM, global simulations may not be able to reproduce the correct spatio-temporal power spectrum of gaseous density fluctuations, and as a result may not properly capture the ISM's gravitational influence on stellar orbits.

In this work, we bridge this gap by analyzing orbital transport driven by a realistic ISM, using the state-of-the-art TIGRESS-NCR magnetohydrodynamics simulations of a local patch of the Galactic disk. In a previous work \citep{Modak2026characterizing}, we developed a direct, field-level, spatio-temporal characterization of realistic ISM density fluctuations, amenable for use in place of cloud-based models in local dynamical calculations. Here, we directly integrate the orbits of ensembles of test particles through the TIGRESS-NCR gravitational fluctuation fields, and measure the resulting orbital transport. We focus on quantifying the amplitude and rates of radial heating and migration induced by the ISM by measuring the evolution of the radial velocity dispersion (or equivalently, the mean radial action) and root-mean-square (rms) changes in guiding center position (or equivalently, the rms changes in angular momentum). We will assess the ISM's contribution to vertical heating and disk thickening in Paper II of this series (S. Modak et al., in preparation). In tandem with the simulation results presented here, we develop a physical interpretation for the unique transport behavior we identify, and show that it agrees with the predictions of quasilinear kinetic theory. We also provide simple fitting formulae for the associated diffusion coefficients, and give prescriptions for how ISM-driven diffusion can be incorporated in (semi-)analytic modeling of stellar dynamics.

The remainder of this paper is organized as follows. In \autoref{sec:methods}, we establish our definitions and notation, and detail the setup of our numerical simulations of ISM-driven transport. In \autoref{sec:solar_neighborhood_sims}, we report and interpret the results of our fiducial solar-neighborhood simulations. In \autoref{sec:discussion}, we assess the role of the vertical structure of the stellar disk on radial transport, compare the dynamical effects of a realistic ISM to those of conventional ISM models, describe how to connect our local results to global galactic models, and compare our findings with observations. We summarize in \autoref{sec:summary}.

\section{Methods}
\label{sec:methods}

In this section, we describe the methods we use to study the influence of ISM fluctuations on stellar orbits. We first review the 3D shearing box approximation and the unperturbed orbits of stars within it (\autoref{sec:shearing_box}). Next, we outline the physical and numerical elements of the TIGRESS-NCR simulations, from which we draw our ISM gravitational potential (\autoref{sec:tigress_simulations}). Finally, we describe the setup we use to integrate test-particle motion in 3D shearing boxes with the ISM potential as an externally-imposed forcing (\autoref{sec:rebound_simulations}), and how we measure the resulting transport from phase space snapshots of the simulation (\autoref{sec:rebound_measurements}).

\subsection{Stellar Orbits in the Shearing Box}
\label{sec:shearing_box}

The ``shearing box'' \citep{GoldreichLyndenBell1965} is a local model of a galactic disk centered at a galactocentric radius $R_0$, and rotating at the circular orbital frequency evaluated at $R_0$, which we denote $\Omega$. The Cartesian coordinates in the box are $(x, y, z) \equiv (R-R_0, R_0[\varphi - \Omega t], z)$, where $(R, \varphi, z)$ are the usual (galactocentric) cylindrical polar coordinates. In this approximation, we assume that the radial extent of the box is small ($|x| \ll R_0$) and that deviations from the box center's circular velocity are small ($|v_x|,\,|v_y| \ll \Omega R_0$), so the background differential rotation of the galaxy can be modeled as a linear shear flow and the epicyclic approximation applies. The equations of motion of a test-particle in the box are
\begin{align}
    \label{eq:shearing_eom_x}
    \ddot{x} & = 2\Omega\dot{y} + 2q\Omega^2 x - \frac{\partial\phi_\mathrm{g}}{\partial x},  \\
    \label{eq:shearing_eom_y}
    \ddot{y} & = -2\Omega\dot{x} - \frac{\partial\phi_\mathrm{g}}{\partial y}, \\
    \label{eq:shearing_eom_z}
    \ddot{z} & = -\frac{\partial\phi_\mathrm{vert}}{\partial z} - \frac{\partial\phi_\mathrm{g}}{\partial z},
\end{align}
where $q \equiv -(\md\ln\Omega/\md\ln R)_{R = R_0}$ parametrizes the shear, $\phi_\mathrm{g}$ is the externally imposed gravitational potential arising from the gas, and $\phi_\mathrm{vert}(z)$ is the vertical potential set by the stellar disk and dark matter halo.\footnote{The contributions of the stars and dark matter to the planar potential are also incorporated into the equations of motion via the choice of azimuthal frequency $\Omega$ and shear parameter $q$.} Note that $\phi_\mathrm{vert}$ remains fixed throughout our simulations, i.e., it is not dynamically sourced by the particles in the simulations we will describe in \autoref{sec:rebound_simulations}. In this notation, Oort's constants are $A \equiv q\Omega/2$ and $B \equiv [(q/2)-1]\Omega$, and a flat rotation curve corresponds to $q = 1$. In the absence of any perturbations, a circular orbit in the midplane ($z \equiv 0$, $v_z \equiv 0$) has a constant radial position $x$, with azimuthal velocity $v_y = -q\Omega x$.

We may regard $\phi_\mathrm{g}$ as a (stochastic) perturbation to the underlying shearing box dynamics. In the unperturbed case (i.e., with $\phi_\mathrm{g} \equiv 0$), \autoref{eq:shearing_eom_x} and \autoref{eq:shearing_eom_y} can be solved to give explicit formulae for the $x$ and $y$ trajectories of test particles (see, e.g., section 8.3.2 of \citealt{BT}):
\begin{align}
    \label{eq:shearing_eom_unperturbed_x}
    x(t) & = x_\mathrm{g}- a_R\cos(\kappa t + \theta_{R,0}), \\
    \label{eq:shearing_eom_unperturbed_y}
    y(t) & = y_0 - q\Omega x_\mathrm{g}t + \gamma a_R \sin(\kappa t + \theta_{R,0}),
\end{align}
where
\begin{equation}
    \label{eq:kappa_definition}
    \kappa \equiv \sqrt{2(2-q)}\Omega
\end{equation}
is the radial epicyclic frequency, $\gamma \equiv 2\Omega/\kappa$, and the constants $x_\mathrm{g}$, $a_R$, $y_0$, and $\theta_{R,0}$ are determined by the particle's initial conditions. Thus, the in-plane component of the unperturbed motion is described by epicyclic oscillations in radius with amplitude $a_R$, initial phase $\theta_{R,0}$, and period $2\pi/\kappa$ around a guiding center position $x_\mathrm{g}$, while the background linear shear flow contributes an azimuthal velocity of $-q\Omega x_\mathrm{g}$ to particles with guiding centers displaced from the box center. Equivalently, this orbit can be described using angle-action variables $(\btheta, \bJ)$, where the actions $\bJ=(J_\varphi, J_R)$ are constants given by
\begin{align}
    \label{eq:Jphi_definition}
    J_\varphi & \equiv \Omega R_0^2 + 2 \Omega R_0 \left(x + \frac{v_y}{2\Omega}\right) \nn \\
    & = \Omega R_0^2 + (2-q)\Omega R_0 x_\mathrm{g}, \\
    \label{eq:JR_definition}
    J_R & \equiv \frac{v_x^2 + \kappa^2(x-x_\mathrm{g})^2}{2\kappa} \nn \\
    & = \frac{1}{2}\kappa a_R^2.
\end{align}
The azimuthal action $J_\varphi$ is the angular momentum relative to the galactic center,\footnote{Often, the azimuthal action is called $J_y$ and is defined without the additive constant $\Omega R_0^2$ so that it is directly proportional to $x_\mathrm{g}$ (see, e.g., \citealt{Fuchs2001}). Within the context of the local shearing box, this constant offset is of no importance, but we choose to include the $\Omega R_0^2$ term to allow a simpler connection between local and global models (e.g., \citealt{GK1,GK2}).} while the radial action $J_R$ measures the extent of the radial epicyclic excursions away from the guiding center. We can relate the mean radial action $\langle J_R \rangle$ of some population of test particles to their rms epicyclic amplitude using \autoref{eq:JR_definition}:
\begin{equation}
    \label{eq:rms_epicyclic_amplitude}
    a \equiv \langle a_R^2\rangle^{1/2} = \left( \frac{2 \langle J_R\rangle}{\kappa}\right)^{1/2}.
\end{equation}
Similarly, if that population is completely phase-mixed (i.e., its distribution function has no $\btheta$-dependence) then using \autoref{eq:JR_definition}, we can relate $\langle J_R \rangle$ to the radial velocity dispersion:
\begin{equation}
    \label{eq:rms_radial_velocity}
    \sigma_R \equiv \sqrt{\langle v_R^2\rangle } =\left(\kappa \langle J_R\rangle \right)^{1/2} = \frac{\kappa a}{\sqrt{2}}.
\end{equation}

Finally, even in the absence of perturbations, the vertical equation of motion, \autoref{eq:shearing_eom_z}, cannot in general be solved explicitly except for very simple choices of $\phi_\mathrm{vert}$. However, for this work we will primarily be interested in orbits with small vertical excursions relative to the scale height of the stellar disk, for which we may expand $\phi_\mathrm{vert}$ about the midplane and only retain terms up to $\mathcal{O}(z^2)$. In this approximation, \autoref{eq:shearing_eom_z} simplifies to
\begin{equation}
    \label{eq:shearing_eom_z_epicycle}
    \ddot{z} \simeq -\nu^2 z - \frac{\partial\phi_\mathrm{g}}{\partial z},
\end{equation}
where the vertical orbital frequency is given by
\begin{equation}
    \label{eq:nu_general}
    \nu \equiv \left(\frac{\partial^2\phi_\mathrm{vert}}{\partial z^2} \bigg|_{z=0}\right)^{1/2}.
\end{equation}
Then, in the absence of perturbations from the gas, orbits take the form
\begin{equation}
    \label{eq:shearing_trajectory_z}
    z(t) = -a_z\cos(\nu t + \theta_{z,0}),
\end{equation}
i.e., oscillations about the midplane with amplitude $a_z$, phase $\theta_{z,0}$, and period $2\pi/\nu$. Analogous to \autoref{eq:JR_definition} above, we can label orbits by their vertical actions
\begin{equation}
    \label{eq:Jz_definition}
    J_z \equiv \frac{v_z^2 + \nu^2 z^2}{2\nu} = \frac{1}{2}\nu a_z^2,
\end{equation}
and analogous to \autoref{eq:rms_radial_velocity} above, for a vertically phase-mixed population we can relate the velocity dispersion to the mean action by
\begin{equation}
    \label{eq:rms_vertical_velocity}
    \sigma_z \equiv \sqrt{\langle v_z^2\rangle } =\left(\nu \langle J_z\rangle \right)^{1/2} = \frac{\nu \langle a_z^2 \rangle^{1/2}}{\sqrt{2}}.
\end{equation}
For typical disk scale heights of $\sim 300\,$pc and vertical periods of $\sim 90\,$Myr, using \autoref{eq:rms_vertical_velocity}, we expect the vertical epicycle approximation to hold when $a_z \lesssim 300\,$pc, or $\sigma_z \lesssim 15\,$km/s for the ensemble. Empirically, we find that this condition is met and the vertical epicycle approximation is reasonably accurate for nearly all ensembles we study in this work, which are most comparable to local thin-disk stars. However, we emphasize that we only use the vertical epicycle approximation to draw initial conditions (see \autoref{sec:rebound_simulations}) and in our interpretation of the simulation results; we do not make any such approximation when running the simulations themselves.

In this paper, we focus only on radial heating (changes in $J_R$) and radial migration (changes in $J_\varphi$)---we defer a detailed analysis of vertical heating (changes in $J_z$) to Paper II of this series.

\subsection{The TIGRESS-NCR Simulations}
\label{sec:tigress_simulations}

To include the effects of a realistic ISM's gravitational potential, we employ a subset of the TIGRESS-NCR simulation suite \citep{Kim2023}. The details of the numerical framework underlying these simulations are described thoroughly in \cite{KimOstriker2017, Gong2017, Kim_JG2017, Kim_JG2023, Kim2023}, but here we briefly review their key features.

The TIGRESS-NCR simulations solve the ideal magnetohydrodynamics (MHD) equations in the shearing box using the \texttt{Athena} code \citep{Stone2008} on a uniform Cartesian mesh, while self-consistently evolving the abundances of atomic, molecular, and ionized hydrogen and explicitly tracking the UV radiation field using adaptive ray tracing. Star formation is modeled by sink particles representing star clusters which form when certain gravitational collapse conditions are met, and can accrete and merge. These star cluster particles are responsible for injecting energy into the gas via a prescription for supernova feedback, as well as sourcing UV photons that can ionize and dissociate the gas, and heat it via a prescription for the photoelectric effect on dust grains. Cosmic-ray heating and ionization is also modeled based on the instantaneous star formation rate. The abundances of key atomic and molecular coolants (C, C$^+$, O, O$^+$, and CO) are derived assuming formation-destruction balance given the local radiation field and hydrogen abundances. Cooling of photoionized gas is due to hydrogen and nebular lines, while cooling of the hot gas assumes collisional and ionization equilibrium. Turbulence is primarily driven by the energy input from supernovae, although the expansion of \ion{H}{2} regions (driven by the pressure of ionized gas) is important to dispersing the dense clouds in which stars form. Gas flows are also modified by Coriolis forces and sheared background rotation, and by gas self-gravity, with the same momentum source terms as appear on the right-hand sides of \autoref{eq:shearing_eom_x} through \autoref{eq:shearing_eom_z}.  The gravitational potential $\phi_\mathrm{g}$ is obtained by solving Poisson's equation via Fourier transforms, and arises from dynamically-evolving structure in the gas density. 

In this paper, we focus primarily on orbits through fluctuations drawn from the ``R8-4\,pc'' (hereafter just ``R8'') simulation, with a nominal Galactocentric radius of $R_0 = 8\,$kpc, a box size of $L_x = L_y \equiv L = 1024\,$pc and $L_z = 6144\,$pc, and a spatial resolution of $\Delta x = 4\,$pc. This simulation is intended to model solar-neighborhood-like ISM conditions, with a mean initial gas surface density of $\overline{\Sigma} = 12\,\Msun/\mathrm{pc}^2$, a circular orbital frequency of $\Omega = 28\,$km/s/kpc, and a flat rotation curve ($q = 1$). The fixed, externally imposed vertical potential due to the stars and dark matter in this simulation is given by
\begin{align}
    \label{eq:external_potential}
    \phi_\mathrm{vert}(z) = 2\pi G \Sigma_* z_* & \left[\left(1 + \frac{z^2}{z_*^2}\right)^{1/2} - 1\right] \nn \\
    & + 2\pi G \rho_\mathrm{DM} R_0^2 \ln\left(1 + \frac{z^2}{R_0^2}\right),
\end{align}
where $\Sigma_* = 42\,\Msun/\mathrm{pc}^2$ is the stellar surface density, $z_* = 245\,$pc is the stellar scale height, and $\rho_\mathrm{DM} = 0.0064\,\Msun/\mathrm{pc}^3$ is the local dark matter volume density \citep{Zhang2013, KimOstriker2017}. For reference, the vertical epicyclic frequency associated with small deviations from the midplane for this potential is given by
\begin{equation}
    \label{eq:nu_TIGRESS}
    \nu = \left(2\pi G\frac{\Sigma_*}{z_*} + 4\pi G \rho_\mathrm{DM}\right)^{1/2},
\end{equation}
which corresponds to a vertical period of $2\pi/\nu \approx 87\,\mathrm{Myr}$ for the R8 parameter values above; note that when contributions from the mean midplane gas potential $\overline{\phi}_\mathrm{g}$ (see below) are included in \autoref{eq:nu_general}, the vertical period decreases by $\sim 10$\,Myr \citep{Kim2020}. In \autoref{sec:LGR4}, we also present the results of orbit integrations using fluctuations from the ``LGR4-2\,pc'' (hereafter just ``LGR4'') simulation, which is representative of ISM conditions with higher mean gas surface densities and star formation rates; see \autoref{sec:LGR4} for a summary of the relevant simulation parameters. Both the R8 and LGR4 simulations reach a statistical steady-state after $\sim 200\,$Myr; in our fiducial orbit integrations described below, we use outputs of the R8 gravitational potential fields at intervals of $1\,$Myr over a temporal baseline $t \in [250\,\mathrm{Myr}, 450\,\mathrm{Myr}]$. For further details on global ISM properties and star formation rates for both the R8 and LGR4 simulations, see \cite{Kim2023}.

From each simulation snapshot, we extract the gas gravitational potential $\phi_\mathrm{g}$, which satisfies Poisson's equation $\nabla^2\phi_\mathrm{g} = 4\pi G \rho$, where $\rho$ is the ISM gas density. The gas gravitational potential can be separated into a (planar-averaged) mean contribution $\overline{\phi}_\mathrm{g}(z, t)$ and a fluctuation $\delta\phi(x, y, z, t) \equiv \phi_\mathrm{g}(x, y, z, t)-\overline{\phi}_\mathrm{g}(z, t)$. The mean gas potential contributes only to the vertical dynamics, while the fluctuations are responsible for driving the in-plane transport. \autoref{fig:orbit_visualization_planar} below shows several snapshots of $\delta\phi$ at the midplane $z = 0$ from the R8 simulation; fluctuations with amplitudes of $\sim (6\, \mathrm{km/s})^2$ are evident. Also notable is that there is more small-scale structure in the negative potential fluctuations (which coincide with dense ISM structures) than in the positive fluctuations.

The mean and fluctuation potentials have been characterized extensively in \cite{Modak2026characterizing} for both the R8 and LGR4 simulations. Briefly, in both simulations, the gravitational potential is well-modeled as inheriting its features (through Poisson's equation) from the surface density field $\Sigma$, coupled with an assumed vertical density profile. In the R8 simulation, at planar spatial wavenumber $|\mathbf{k}|=k \gtrsim 0.01\,\mathrm{pc}^{-1}$, the power spectrum of surface density fluctuations $\delta \equiv (\Sigma - \overline{\Sigma})/\overline{\Sigma}$ drops as a steep power law,
\begin{equation}
    \label{eq:Pdelta_scaling}
    P_\delta(k) \propto k^{-n_\delta},
\end{equation}
where $n_\delta = 2.3$, while becoming slightly shallower at smaller $k$. The LGR4 simulations also exhibit a similarly steep spatial power spectrum. Thus, the power is dominated by fluctuations at small values of $k$ (large scales). Surface density fluctuations in both simulations are also \textit{temporally} correlated, with structure lifetimes at the largest spatial scales set primarily by feedback-driven-dispersal. In \autoref{sec:GRF_model}, we will compare the transport that results from the full gravitational potential extracted from the R8 simulation to the transport resulting from a Gaussian random field (GRF) with a matching spatio-temporal power spectrum. We refer readers interested in further details of properties of the emergent gravitational potential from the TIGRESS-NCR simulations to Section 4 of \cite{Modak2026characterizing}.

\subsection{Numerical Orbit Integrations using \texttt{REBOUND}}
\label{sec:rebound_simulations}

In the presence of the perturbing gravitational potential of the ISM, the actions given by \autoref{eq:Jphi_definition} and \autoref{eq:JR_definition} will in general not remain constant, and stars will be gradually ``transported'' across action space. The main purpose of this paper is to measure and understand ``in-plane'' transport processes, namely radial heating and radial migration; we will evaluate the ISM's role in driving vertical transport in Paper II. To do this, we perform a series of test-particle shearing box simulations using the \texttt{REBOUND} code \citep{ReinLiu2012}, and subject the particles (hereafter ``stars'') to external forcing determined by the TIGRESS-NCR ISM potential.

We configure the \texttt{REBOUND} shearing box to match the parameters $\{L, \Omega, q\}$ of the \texttt{Athena} shearing box used in the TIGRESS-NCR simulations from which we draw the ISM potential (see \autoref{sec:tigress_simulations}), aligning the time coordinate so that the same ``shearing-periodic'' boundary conditions \citep{HawleyGammieBalbus1995} used in the TIGRESS-NCR simulations apply to the stars consistently. We integrate the equations of motion (\autoref{eq:shearing_eom_x} through \autoref{eq:shearing_eom_z}) using the symplectic epicycle integrator (SEI) implemented in \texttt{REBOUND} \citep{ReinTremaine2011}, except we modify it to:
\begin{enumerate}[label=(\roman*)]
    \item account for the difference in shear profiles between the Keplerian ($\kappa = \Omega$, $q = 3/2$) and flat rotation curve ($\kappa = \sqrt{2}\Omega$, $q = 1$) contexts,
    \item integrate vertical dynamics in $\phi_\mathrm{vert}$ (see \autoref{eq:external_potential}) instead of the default harmonic vertical potential, and
    \item incorporate the perturbing potential of the ISM read in from a TIGRESS-NCR simulation snapshot as an external force.
\end{enumerate}
We provide a detailed description of our \texttt{REBOUND} integration routine in \autoref{sec:REBOUND_details}, but briefly, (i) we add a new constant $q$ and modify the integrator update rules to allow for arbitrary shear profiles, (ii) we integrate the vertical motion as a sub-stepping process within the SEI, and (iii) we calculate the accelerations $-\nabla\phi_\mathrm{g}$ using second-order finite differences on the TIGRESS-NCR spatial grid, with linear interpolation in time between TIGRESS-NCR snapshots. We integrate orbits for 10\,Gyr ($\approx 45\times 2\pi/\Omega$) using a timestep of $\Delta t = 10^{-4}\times 2\pi/\Omega$ ($\approx 0.02\,$Myr).\footnote{Note that because the temporal baseline of TIGRESS-NCR snapshots is 200\,Myr, this integration involves looping the sequence of snapshots 50 times, which is justifiable because the ISM is in statistical steady-state over all snapshots used in this analysis (e.g., \citealt{Kim2023}), and the correlation time of the dominant fluctuations is smaller than 200\,Myr. The precise looping procedure is described in \autoref{sec:REBOUND_details}; we have checked that it does not introduce any numerical artifacts and that our results are insensitive to the looping period (i.e., if instead we loop the first 100\,Myr of the snapshots 100 times, our results are unchanged).} With this timestep choice, we find relative energy errors of at most $|\Delta E/E| \sim 10^{-6}$ in the absence of external fluctuations, and have checked that our integrations are converged by ensuring that using a $3\times$ smaller timestep does not materially alter our results.

\begin{figure*}
    \centering
    \includegraphics[width=0.99\textwidth]{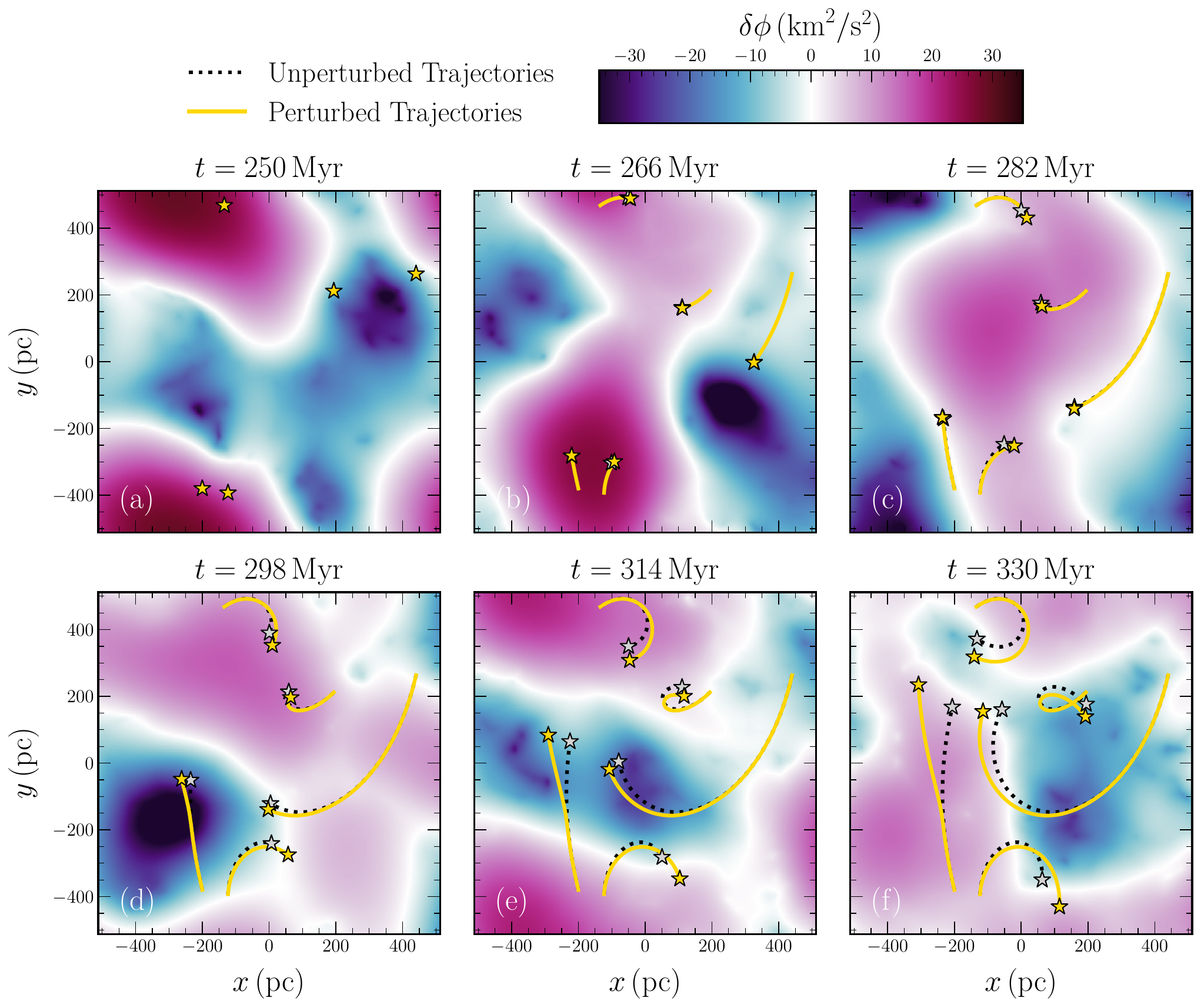}
    \caption{Snapshots of the gravitational potential fluctuations extracted from the TIGRESS-NCR R8 simulation at $z = 0$, with five example trajectories simulated using the method described in \autoref{sec:rebound_simulations} overlaid. The gold stars indicate the stellar positions at the snapshot time, and the gold curves show the stars' trajectories starting from the initial conditions shown in panel (a) as they evolve according to \autoref{eq:shearing_eom_x} through \autoref{eq:shearing_eom_z} for approximately half an epicyclic period $\pi/\kappa \approx 80\,$Myr. For comparison, we plot the unperturbed trajectories (i.e., solutions using the same initial conditions as panel (a) but with $\phi_\mathrm{g} \equiv 0$) as black dotted curves, and indicate those positions at each snapshot time with grey stars.}
    \label{fig:orbit_visualization_planar}
\end{figure*}

In \autoref{fig:orbit_visualization_planar}, we illustrate the effects of the ISM fluctuations on a few representative stellar orbits: the yellow stars indicate the stellar positions at the same snapshot time as the TIGRESS-NCR R8 gravitational potential fluctuations, while the yellow trails show the trajectory they trace out in the presence of the fluctuations over half a radial orbital period ($\pi/\kappa \approx 80\,$Myr). The black dotted curves show orbits with the same initial conditions for the same duration, but in the absence of the ISM fluctuations, to illustrate the effect of the perturbations in altering the orbit's phase and actions.

In what follows, we will analyze the results of several simulations of $N = 50,000$ of these stellar orbits each to study the ensemble-averaged, secular effects of the perturbations visualized in \autoref{fig:orbit_visualization_planar}. Across all of our simulations, we initialize stars with a uniform distribution in planar angles, i.e., we draw values of $\theta_{R,0} \in [0, 2\pi)$ and $y_0 \in [-L/2, L/2)$ uniformly. For the azimuthal actions, we draw local guiding centers $x_\mathrm{g}\in [-L/2, L/2)$ uniformly, and then apply \autoref{eq:Jphi_definition}. For the radial actions, we sample from a Schwarzschild distribution function (DF) with mean radial action $\langle J_R \rangle_0$,
\begin{equation}
    \label{eq:schwarzschild_IC}
    f(J_R) = \frac{1}{\langle J_R \rangle_0} e^{-J_R/\langle J_R \rangle_0},
\end{equation}
which we parametrize by the initial radial velocity dispersion $\sigma_{R,0} = \left(\kappa\langle J_R \rangle_0)\right)^{1/2}$. Finally, for the vertical initial conditions, we assume the epicycle approximation given by \autoref{eq:shearing_eom_z_epicycle}, and derive initial values for vertical position and velocity using \autoref{eq:shearing_trajectory_z}. We initialize stars with a uniform distribution in vertical angle $\theta_{z,0} \in [0, 2\pi)$, and sample vertical actions from a DF analogous to \autoref{eq:schwarzschild_IC} of the form
\begin{equation}
    \label{eq:schwarzschild_IC_z}
    f(J_z) = \frac{1}{\langle J_z \rangle_0} e^{-J_z/\langle J_z \rangle_0},
\end{equation}
which we parametrize by the initial vertical velocity dispersion $\sigma_{z,0} = (\nu\langle J_z \rangle_0)^{1/2}$. For reference, this scheme of sampling produces an equilibrium DF that will not evolve in the absence of fluctuations,\footnote{Strictly, such a procedure only yields a vertically phase-mixed distribution to the extent that the vertical epicycle approximation holds, i.e., when $\sigma_{z, 0} \lesssim 15\,$km/s, which is true for all fiducial simulations we describe in \autoref{sec:solar_neighborhood_sims} below (though see \autoref{fig:vertical_suppression} and Footnote 14).} and corresponds to a product of Maxwellian distributions in velocity space, with a dispersion of $\sigma_{R,0}$ in $v_x$, the radial velocity, a dispersion of $\sigma_{R,0}/\gamma$ in the quantity
\begin{equation}
    \label{eq:vytilde_definition}
    \tilde{v}_y \equiv v_y + q\Omega x,
\end{equation}
the azimuthal velocity relative to the local circular orbit, and a dispersion of $\sigma_{z,0}$ in $v_z$, the vertical velocity. As we describe in \autoref{sec:solar_neighborhood_sims}, for much of this paper we will focus on ensembles that are initially cold in the vertical direction ($\sigma_{z,0} = 0$). However, we do analyze the effect of nonzero initial vertical dispersion ($\sigma_{z,0} \neq 0$) on radial heating and migration in \autoref{sec:vertical_decoupling}. We will thoroughly assess vertical heating in Paper II.

\subsection{Measuring Radial Transport from \texttt{REBOUND} Snapshots}
\label{sec:rebound_measurements}

There are several ways to measure radial transport in simulations: the most familiar diagnostic is to measure radial heating via the standard deviation of radial velocities, $\sigma_R$, of the stars in each snapshot. In addition to this, we will analyze the time-evolution of radial and azimuthal actions for individual stars, as follows.

Using \autoref{eq:shearing_eom_unperturbed_x}, \autoref{eq:shearing_eom_unperturbed_y}, and \autoref{eq:vytilde_definition}, and given the values of a star's position and velocity within the box, we can identify its guiding center position as
\begin{equation}
    \label{eq:xg_vtilde}
    x_\mathrm{g}= x + \frac{\tilde{v}_y}{(2-q)\Omega}.
\end{equation}
Together with \autoref{eq:JR_definition}, this also allows us to quantify radial heating by calculating the change in radial action $\delta J_R(t) \equiv J_R(t) - J_R(0)$ of each star as a function of time. Note that \autoref{eq:xg_vtilde} can yield a value of $x_\mathrm{g}$ outside the box, but in practice we can always add or subtract $L$ from the result until $-L/2 \leq x_\mathrm{g}< L/2$, as long as the same shifts are applied to $x$. These shifts do not alter measurements of the radial action, since $J_R$ only depends on the \textit{difference} $(x-x_\mathrm{g})$.

However, the azimuthal action $J_\varphi$ depends not just on the difference $(x-x_\mathrm{g})$, but on the value of $x_\mathrm{g}$ itself. As a result, na\"ively applying \autoref{eq:xg_vtilde} in conjunction with \autoref{eq:Jphi_definition} can lead to spurious results, since the quantity $x$ is measured in a shearing-periodic box: effectively, stellar trajectories will not ``remember'' having crossed the box boundary, which instantaneously shifts the value of $x$ by $\pm L_x$. In order to account for these boundary crossings, we take the time derivative of \autoref{eq:xg_vtilde} to obtain the rate of change of the guiding center,
\begin{equation}
    \frac{\md x_\mathrm{g}}{\md t} = v_x + \frac{1}{(2-q)\Omega}\frac{\md \tilde{v}_y}{\md t},
\end{equation}
which has the advantage that both $v_x$ and $\tilde{v}_y$ remain unchanged when a stellar trajectory crosses the box boundaries. Thus, we can calculate the amount each star has migrated, $\delta x_\mathrm{g}(t) \equiv x_\mathrm{g}(t)-x_\mathrm{g}(0)$, by numerically integrating values from stored phase space snapshots:
\begin{equation}
    \delta x_\mathrm{g}(t) = \int_0^t v_x(t') \md t' + \frac{\delta\tilde{v}_y(t)}{(2-q)\Omega},
\end{equation}
where $\delta \tilde{v}_y(t) \equiv \tilde{v}_y(t) - \tilde{v}_y(0)$. We then apply \autoref{eq:Jphi_definition} to calculate $\delta J_\varphi(t) \equiv J_\varphi(t) - J_\varphi(0) = (2-q)\Omega R_0 \delta x_\mathrm{g}$. The snapshots we use to analyze radial migration are taken every $2\,$Myr ($\approx 10^{-2}\times 2\pi/\Omega$); we have checked that using more frequent snapshots does not materially alter our measured migration, and we are able to reliably track migration for all stars over scales larger than the the box itself as needed (see, e.g., \autoref{fig:deltaxg_hist_t} below).

\section{Solar-Neighborhood Simulations}
\label{sec:solar_neighborhood_sims}

In this section, we present the results of our fiducial simulations following the method described in \autoref{sec:rebound_simulations}. For these simulations, we draw ISM fluctuation fields from the the TIGRESS-NCR R8 model, which is designed to match the characteristics of the present-day solar neighborhood (see \autoref{sec:tigress_simulations}). However, we have also run all such simulations under the alternative LGR4 model, which is intended to model ISM conditions at higher gas surface densities. We give details of the LGR4 runs and discuss how they are related to our R8 results in \autoref{sec:LGR4}.

Throughout this section, we also assume for simplicity that stars are initialized in-plane with zero initial vertical velocity,\footnote{We re-emphasize that this is a choice concerning initial conditions only: we do include vertical forcing throughout our simulations, so stars are \textit{not} permanently confined to the midplane.} i.e., $\sigma_{z,0} = 0$. This is not particularly restrictive, because it turns out that radial transport is largely decoupled from vertical motion---in \autoref{sec:vertical_decoupling}, we demonstrate that choosing $\sigma_{z,0} > 0$ leads only to a weak, quantifiable suppression in the \textit{amplitude} of radial transport without any alteration to the transport \textit{scalings} themselves. Having chosen $\sigma_{z,0} = 0$, the only parameter governing the initial conditions is the initial radial velocity dispersion $\sigma_{R,0}$.

We describe the key results of our fiducial simulations in \autoref{sec:results}, interpret them in terms of quasilinear diffusion theory in \autoref{sec:physical_interpretation}, and present measurements of action-space diffusion coefficients that quantify ISM-driven radial transport in \autoref{sec:diffusion_measurements}. Additionally, in \autoref{sec:GRF_model} we compare our findings to simulations involving fluctuations produced using a Gaussian random field (GRF) model for the ISM surface density instead of the true TIGRESS-NCR fluctuations.

\subsection{Fiducial Simulation Results}
\label{sec:results}

\begin{figure}
    \centering
    \includegraphics[width=0.47\textwidth]{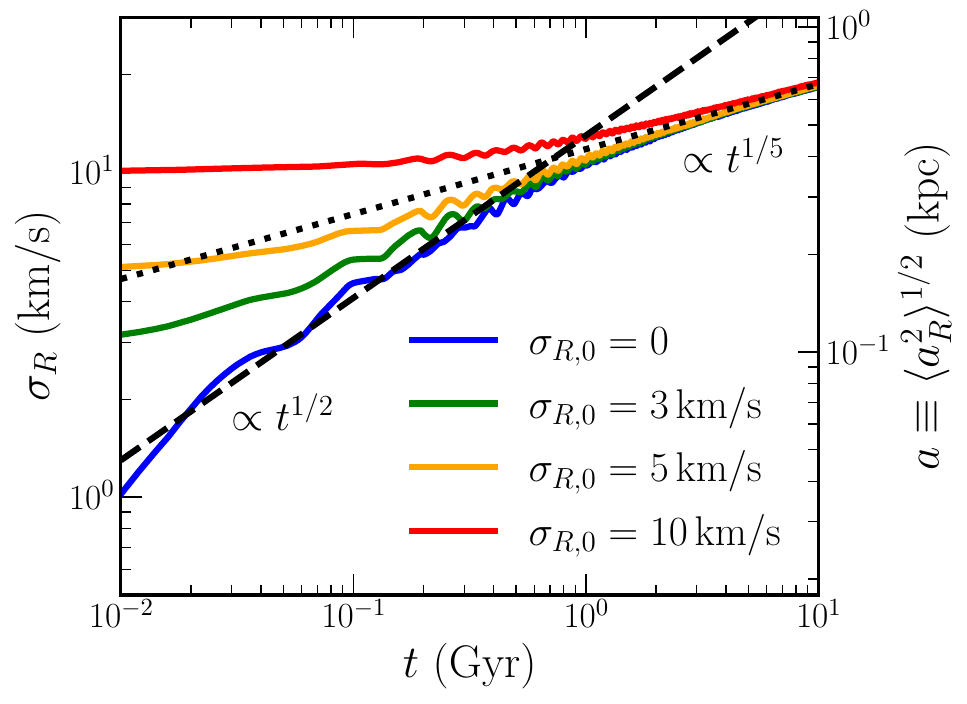}
    \caption{The radial velocity dispersion of the stars as a function of time for our fiducial (solar-neighborhood-like) simulations with varied initial radial velocity dispersions (different colors). The black dashed and dotted lines indicate the scalings $t^{1/2}$ and $t^{1/5}$ respectively, which are the predicted scalings in the long- and short-wavelength regimes respectively (see \autoref{eq:long_wavelength_scalings} and \autoref{eq:short_wavelength_scalings} in \autoref{sec:physical_interpretation}). For reference, the ensemble's rms epicyclic amplitude $a = \sqrt{2}\sigma_R/\kappa$ (see \autoref{eq:rms_epicyclic_amplitude}) is indicated on the right axis.}
    \label{fig:sigmaR_t}
\end{figure}

In \autoref{fig:sigmaR_t} we plot the radial velocity dispersion of stars, $\sigma_R$, as a function of time for four ensembles, labeled by their initial radial dispersion $\sigma_{R,0}$. We see that for the initially completely cold ensemble (blue curve), the ensemble's radial velocity dispersion follows roughly a broken power-law scaling, with $\sigma_R\propto t^{1/2}$ at early times and $\sigma_R\propto t^{1/5}$ at late times. The transition between these regimes begins roughly when $\sigma_R \approx 6$\,km/s, corresponding to an rms epicylic amplitude $a\approx 210$\,pc. For all other ensembles, the scaling $\sigma_R \propto t^{1/5}$ is visible at late times, although there is not a particularly well-defined scaling at the earlier times. In all cases, the final radial velocity dispersion after $10$\,Gyr is $\sigma_R\approx 18\,$km/s.

\begin{figure}
    \centering
    \includegraphics[width=0.45\textwidth]{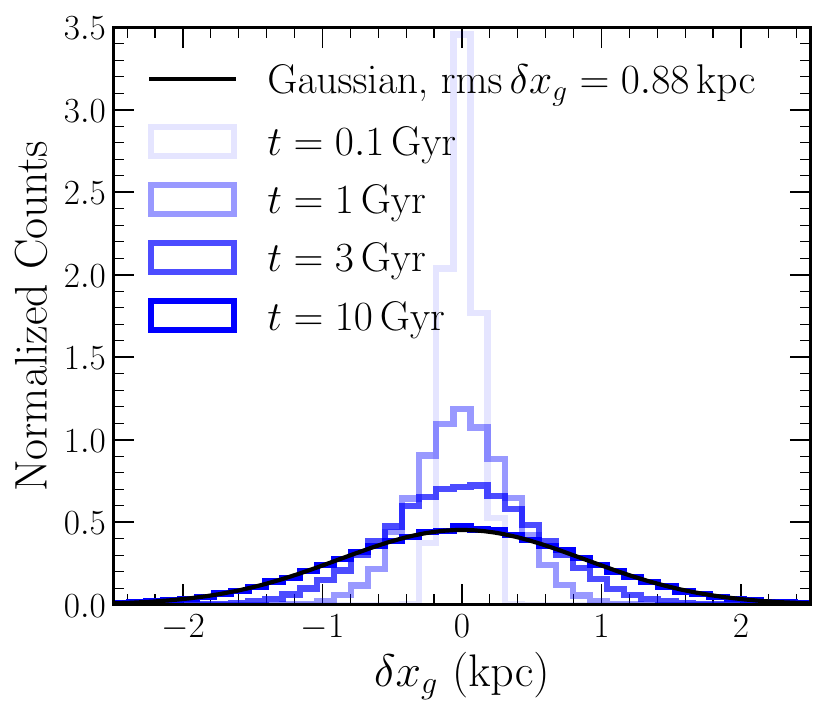}
    \caption{The distribution of changes in stellar guiding center positions $\delta x_\mathrm{g}$ as a function of time (indicated by color, progressing from light to dark) for the initially cold ($\sigma_{R,0} = 0$) fiducial simulation. The black curve is a Gaussian fit with the same total measured rms change in guiding center as the final simulation snapshot.}
    \label{fig:deltaxg_hist_t}
\end{figure}

In \autoref{fig:deltaxg_hist_t}, we plot the histogram of the guiding center changes $\delta x_\mathrm{g}$ of all stars in the initially cold ($\sigma_{R,0}=0$) ensemble at various times $t$. To a good approximation, the histogram spreads as a Gaussian centered on $\delta x_\mathrm{g}=0$, characteristic of a diffusion process with an approximately constant diffusion coefficient. In particular, the final histogram at $t=10$\,Gyr is well-fit by a Gaussian with standard deviation $\mathrm{rms}\,\delta x_\mathrm{g} \approx 0.88$\,kpc. Though we do not show them here, the other ensembles with varied $\sigma_{R,0}$ show near-identical behavior (see panel (b) of \autoref{fig:rmsdJ_t} below).

\begin{figure}
    \centering
    \includegraphics[width=0.47\textwidth]{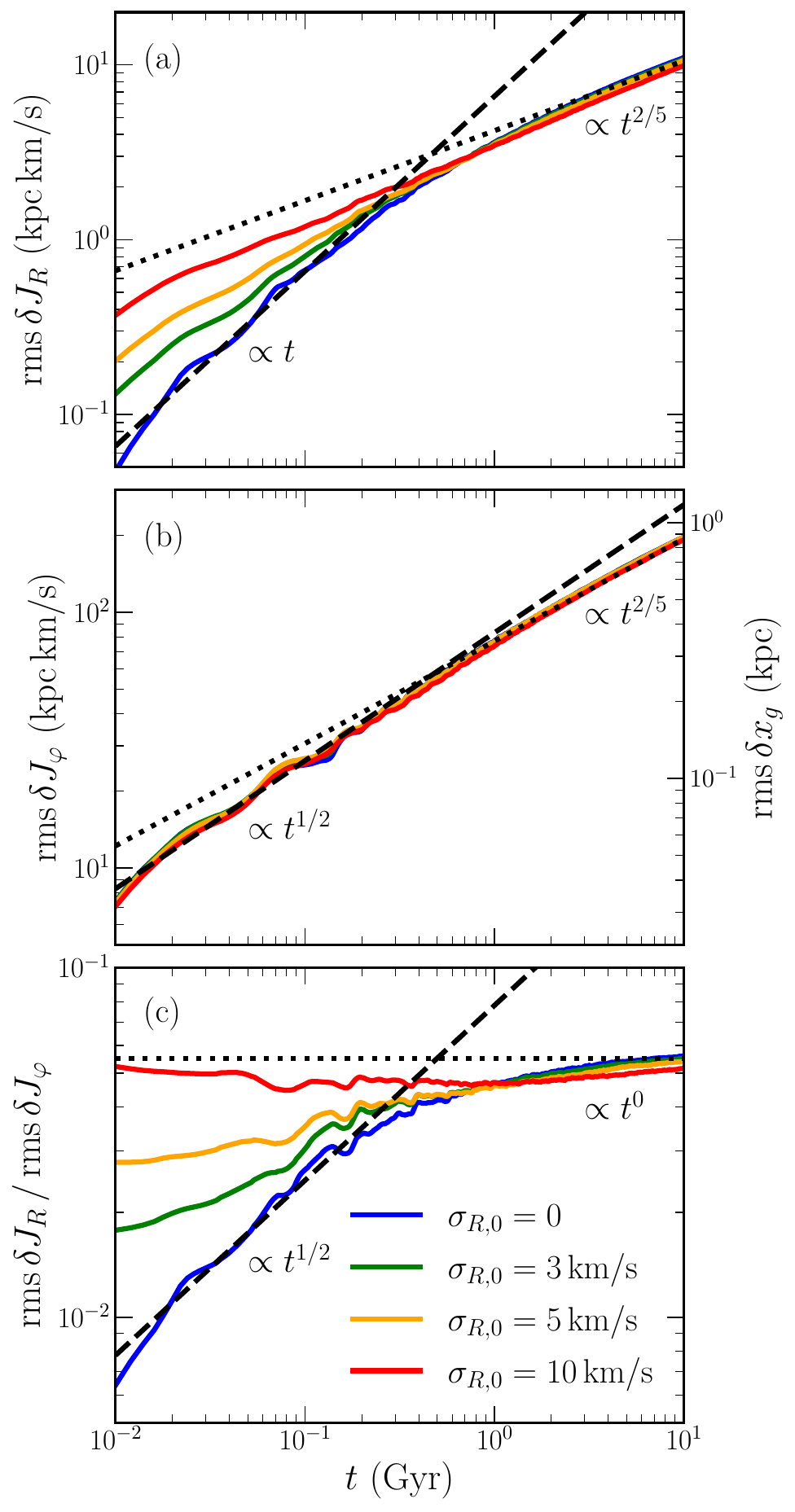}
        \caption{The rms changes in stellar radial (panel (a)) and azimuthal (panel (b)) actions (see \autoref{eq:rmsdJ_definition}), and their ratio (panel (c)), for the fiducial simulations with varied initial velocity dispersions (different colors). Black dashed and dotted lines indicate the scalings predicted by theory in the long- and short-wavelength regimes respectively (see \autoref{eq:long_wavelength_scalings} and \autoref{eq:short_wavelength_scalings} in \autoref{sec:physical_interpretation}). In panel (b), we also indicate the corresponding rms changes in guiding center positions on the right axis.}
        \label{fig:rmsdJ_t}
\end{figure}

To build a better physical understanding of the radial transport behavior shown in \autoref{fig:sigmaR_t} and \autoref{fig:deltaxg_hist_t}, we convert these results into action space. In \autoref{fig:rmsdJ_t} we plot the rms action deviations, rms\,$\delta J_R$ and rms\,$\delta J_\varphi$, defined by
\begin{equation}
    \label{eq:rmsdJ_definition}
    \mathrm{rms}\,\delta J_{i}(t) \equiv \left(\frac{1}{N}\sum_{j=1}^{N} \left[J_i^{(j)}(t) - J_i^{(j)}(0)\right]^2\right)^{1/2},
\end{equation}
where $i = R, \varphi$ and the index $j$ runs over the $N$ stars in the ensemble. The scalings in panel (a) are related straightforwardly to those in \autoref{fig:sigmaR_t} following \autoref{eq:rms_radial_velocity}, because we have rms\,$\delta J_R \simeq \sqrt{2}\langle J_R \rangle = \sqrt{2} \sigma_R^2 / \kappa$, where in the first approximate equality we have assumed a Schwarzschild DF as given in \autoref{eq:schwarzschild_IC}, and that the initial radial actions are small (see also the discussion in \autoref{sec:rms_from_D}). Meanwhile, in panel (b), the quantity rms\,$\delta J_\varphi = (2-q)\Omega R_0 \, \mathrm{rms}\,\delta x_\mathrm{g}$ exhibits the scaling $\propto t^{1/2}$ at early times and $\propto t^{2/5}$ at late times; the turnover between these two scalings occurs at around the same time as the turnover exhibited by the blue line in panel (a). Finally, in panel (c) we plot the radial heating-to-migration ratio, rms\,$\delta J_R$\,/\,rms\,$\delta J_\varphi$. Again the coldest (blue) ensemble has distinct early- and late-time scalings---with rms\,$\delta J_R$\,/\,rms\,$\delta J_\varphi \propto t^{1/2}$ at early times and roughly independent of $t$ at late times---while the other ensembles only clearly exhibit the late-time scaling. Note that for all ensembles, the ratio after $10$\,Gyr converges to rms\,$\delta J_R$\,/\,rms\,$\delta J_\varphi \approx 0.055$.

\begin{figure}
    \centering
    \includegraphics[width=0.47\textwidth]{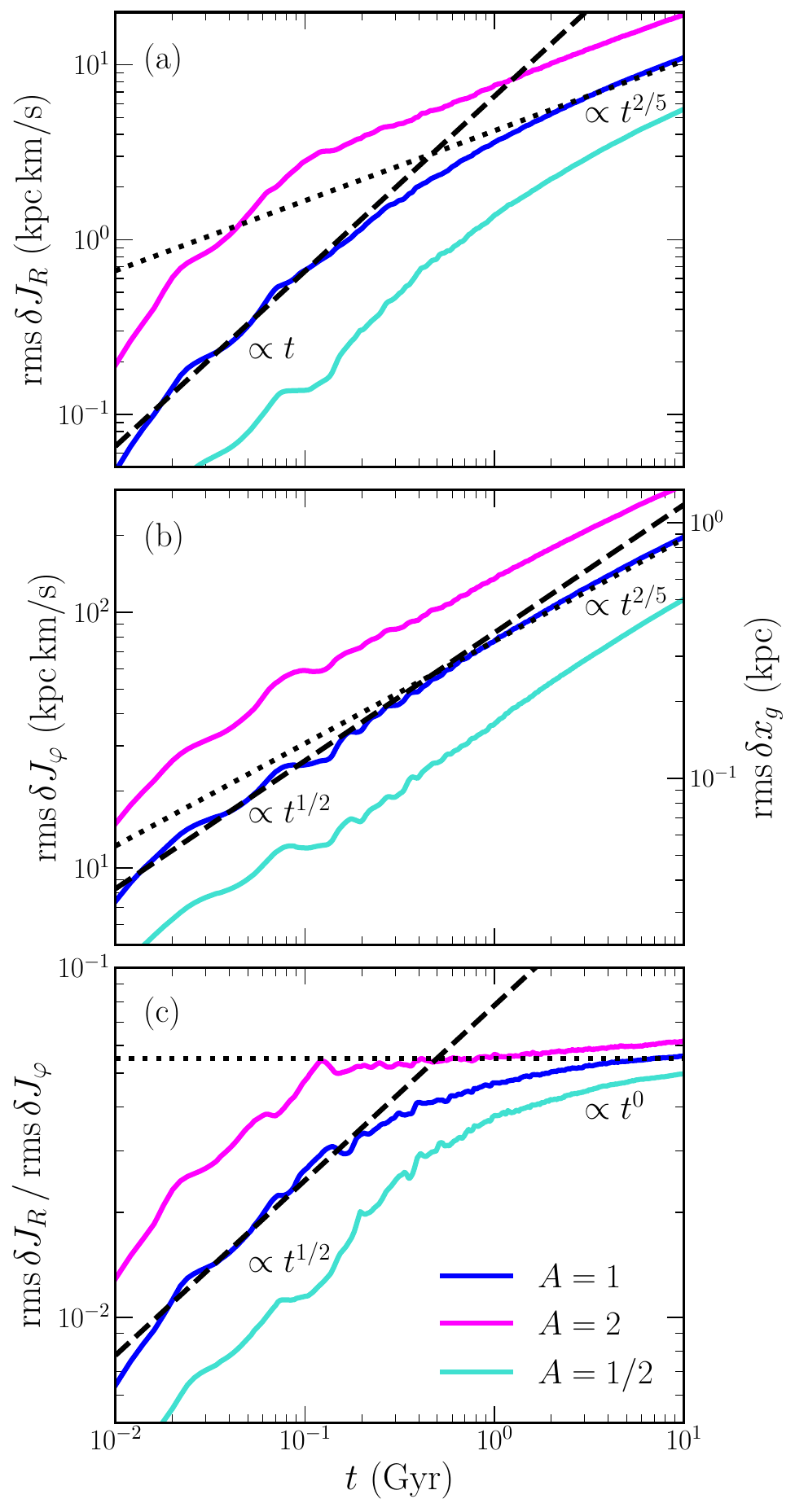}
    \caption{As in \autoref{fig:rmsdJ_t}, except only for initially cold ($\sigma_{R,0} = 0$) ensembles, integrated in ISM fluctuations with amplitudes multiplied by an overall factor of $A = 1/2$ (cyan) or $2$ (magenta). For comparison, we also show the ensemble integrated in the unaltered ($A = 1$) fluctuations (blue, same as the blue curves in \autoref{fig:rmsdJ_t}).}
    \label{fig:rmsdJ_t_variedA}
\end{figure}

Although the R8 simulation is calibrated on the solar-neighborhood conditions today, it is likely that the gas density fluctuations in the solar-neighborhood had significantly different amplitudes in the past, as the mean gas surface density was larger (see, e.g., Figure 5 of \citealt{TacconiGenzelSternberg2020}). To investigate the role played by this amplitude, we ran some further simulations assuming perfectly cold initial ensembles ($\sigma_{R,0}=0$) but multiplying the ISM fluctuation fields by a factor $A$. In \autoref{fig:rmsdJ_t_variedA} we show the heating and migration results for $A = 1/2$, $1$, and $2$. The blue $A = 1$ curves are identical to those in \autoref{fig:rmsdJ_t}. The other curves exhibit the same basic power law scalings at early and late times, but the transition between the power-law regimes occurs earlier (later) for the simulation with stronger (weaker) forcing. Similarly, the overall amplitude of measured transport is larger (smaller) when we apply stronger (weaker) forcing, although the heating-to-migration ratio approaches approximately the same value at late times.

\subsection{Interpretation as a Diffusion Process}
\label{sec:physical_interpretation}

The results shown in \autoref{fig:sigmaR_t} through \autoref{fig:rmsdJ_t} can be explained relatively simply using the formalism of (quasilinear) diffusion theory. Roughly speaking, quasilinear diffusion theory is applicable in this context because ISM fluctuations are (i) weak compared to the mean-field gravitational forces produced by the stellar disk and dark matter halo, and (ii) short-lived compared to the time over which the stars' orbital actions change significantly. The detailed justification for and application of this theory will be given in an upcoming paper (C. Hamilton et al., in preparation; see also Section VII of \citealt{HamiltonFouvry2024}); below, we just summarize some of the key aspects necessary for interpreting our results.

In quasilinear theory, the angle-averaged distribution function of stellar orbits $f_0(\bJ, t)$ evolves via a diffusion equation,
\begin{equation}
    \label{eq:quasilinear_diffusion}
    \frac{\partial f_0}{\partial t} = \frac{1}{2}\sum_{i}\sum_{j} \frac{\partial}{\partial J_i}\left(D_{ij}(\bJ)\frac{\partial f_0}{\partial J_j}\right),
\end{equation}
where $i, j = R, \varphi$ here and below. Note that the evolution is purely diffusive with no frictional drag, because we are treating the potential fluctuations as externally imposed rather than self-consistently generated by (or at all responsive to) fluctuations in the stellar distribution.\footnote{Of course, in principle the ISM \textit{will} contribute to a drag $\mathbf{a}_{\mathrm{DF}}$ on each star arising from dynamical friction. A simple estimate of this drag \citep{Ostriker1999} evaluated along our simulated orbits yields $|\mathbf{a}_{\mathrm{DF}}|/|\nabla\phi_\mathrm{g}| \lesssim 10^{-3}$, which is negligible.} Assuming that the diffusion tensor $D$ varies only weakly with $J_\varphi$ (as is appropriate given the local approximation underlying our shearing box), and that the DF takes the Schwarzschild form (see \autoref{eq:schwarzschild_IC}) with $\langle J_R \rangle /  J_\varphi \sim \sigma_R^2/(\kappa\Omega R^2) \ll 1$, it is easy to relate the radial transport behavior we are studying here to the diagonal\footnote{In \autoref{sec:rms_from_D}, we show that the off-diagonal components $D_{\varphi R} = D_{R\varphi}$ do not contribute to radial heating in the shearing box, and are typically subdominant in contributing to radial migration as well, so we neglect them here for simplicity.} components of the diffusion tensor $D_{RR}$ and $D_{\varphi\varphi}$. In particular, we show in \autoref{sec:rms_from_D} that if the  coefficients in the diffusion tensor have power-law dependence on $J_R$,
\begin{align}
    \label{eq:DRR_power}
    D_{RR}(\bJ) & \equiv d_{RR} J_R^{\alpha_R}, \\
    \label{eq:Dphiphi_power}
    D_{\varphi\varphi}(\bJ) & \equiv d_{\varphi\varphi} J_R^{\alpha_\varphi},
\end{align}
for some coefficients $d_{RR}, d_{\varphi\varphi}$ and power law indices $\alpha_R\neq 2, \alpha_\varphi$, then the radial heating and migration approximately satisfy
\begin{align}
    \label{eq:rms_dJR_approx}
    \mathrm{rms}\,\delta J_R(t) & \simeq \sqrt{2} \, c(\alpha_R) \times (d_{RR} t)^{1/(2-\alpha_R)}, \\
    \label{eq:rms_dJphi_approx}
    \mathrm{rms}\,\delta J_\varphi(t) & \simeq \left[\frac{(2-\alpha_R)\Gamma(1+\alpha_\varphi)}{2-\alpha_R + \alpha_\varphi} c(\alpha_R)^{\alpha_\varphi}\right]^{1/2} \nn \\
    & \quad\quad \times d_{\varphi\varphi}^{\,1/2} \, d_{RR}^{\,\alpha_\varphi/[2(2-\alpha_R)]} \, t^{1/2 + \alpha_\varphi/[2(2-\alpha_R)]},
\end{align}
respectively, where $\Gamma(z)$ is the Gamma function and $c(\alpha_R)$ is a function of the power law index $\alpha_R$ alone, defined in \autoref{eq:c_alphaR_definition}. Thus, if we can understand the $J_R$-dependence of the diffusion coefficients $D_{RR}$ and $D_{\varphi\varphi}$, we should be able to understand the power-law scalings exhibited in the simulations.

The quasilinear diffusion coefficients $D_{ij}$ are determined by the spatio-temporal power spectrum of the potential fluctuations, $P_{\delta\phi}$: following, e.g., Equation 3.9a of \cite{BinneyLacey1988}, we have
\begin{equation}
    \label{eq:D_from_power}
    D_{ij}(\bJ) = \frac{1}{2}\sum_{n_\varphi} \sum_{n_R} n_i n_j \tilde{P}_{\delta\phi}(n_\varphi, n_R, \bJ, \omega_\mathrm{res}),
\end{equation}
where $n_\varphi, n_R \in \mathbb{Z}$ are the Fourier ``wavenumbers'' associated with the angles $\theta_\varphi$ and $\theta_R$ respectively. In this expression, $\tilde{P}_{\delta\phi}$ is the component of the spatio-temporal spectrum associated with a given $(n_\varphi, n_R)$ pair, evaluated at the action $\bJ$ and frequency\footnote{In general, the spectrum should be evaluated at the frequency $\omega = n_R\kappa + n_\varphi(\Omega_\varphi - \Omega)$, where $\Omega_\varphi$ is the azimuthal frequency and we have subtracted $\Omega$ because we are working in the rotating frame of the box. For the shearing box approximation, $\Omega_\varphi - \Omega = -q\Omega x_\mathrm{g}/R_0$.} $\omega_{\mathrm{res}} \equiv n_R\kappa - n_\varphi (q\Omega x_\mathrm{g}/R_0)$. Expressions for translating potential fluctuations (and hence their power spectra) to angle-action coordinates for the epicyclic orbits we consider here, and subsequently Fourier transforming in angles, are given in Section 4 and Appendix B of \cite{GK1}. The spatio-temporal spectra of ISM potential fluctuations in the TIGRESS-NCR R8 and LGR4 simulations have been measured and modeled by \cite{Modak2026characterizing}; we highlight their salient features pertaining to radial transport below.

Radial transport is dominated by ISM fluctuations at a particular \textit{wavelength}, $\lambda_*$, or equivalently a special wavenumber\footnote{Strictly, these fluctuations are slightly anisotropic, with a preference for trailing structures (see, e.g., Figure 6 of \citealt{Modak2026characterizing}), but we find that a simplified, isotropic fluctuation model suffices for building a qualitative understanding of the transport (see \autoref{sec:GRF_model}), and so refer to scalar wavenumbers rather than vectors throughout this analysis.} $k_*\equiv 2\pi/\lambda_*$. This dominance occurs for two reasons. First, for most scales $k$ of relevance to the ISM, the spatial power spectrum exhibits a steep power-law behavior (see \autoref{eq:Pdelta_scaling}). Small scales (high $k$) contribute very little power to the total potential fluctuations, and so drive negligible heating and migration---we confirm this result explicitly in \autoref{sec:comparison_conventional}. Secondly, at the large-scale (low $k$) end of this spectrum, the steep power law does not continue forever. Instead, it turns over at a wavenumber $k_*$  (typically $\sim 0.01\,\mathrm{pc}^{-1}$ in the solar neighborhood) corresponding to the scale $\lambda_*$ (typically $2\pi/k_* \sim 600\,$pc in the solar neighborhood). We do believe that this dominant scale is physical (measuring the distance over which supernovae feedback is correlated) rather than artificial (set by the R8 simulation's box size), and have tested this by measuring the dominant scale in some lower-resolution TIGRESS simulations
of the solar neighborhood with larger box sizes, finding very similar values of $\lambda_* \sim 600\,\mathrm{pc}$. However, we have not rigorously tested whether the same dominant scale applies across many different galactic environments (see also \autoref{sec:environmental_dependence}). and so cannot claim here that it applies universally. In any case, the upshot is that in the solar neighborhood, ISM-driven radial transport is dominated by fluctuations on a scale $\lambda_* \sim 600\,$pc.

There is also a special timescale in our problem, which is the typical correlation time of fluctuations with wavenumber $k_*$. As shown in Figure 9 of \cite{Modak2026characterizing}, 
for a given $k$ the spatio-temporal power spectrum (measured in the frame rotating with the center of our shearing box) is relatively flat at low frequencies, then drops off sharply at high frequencies. The transition between these two behaviors occurs around $\omega_c(k) \approx 0.4 \omega_0(k)$, where $\omega_0(k) = [\tau_0^{-2} + v_\mathrm{eff}^2 k^2]^{1/2}$ and the measured parameter values in the R8 simulation are $\tau_0 = 5$\,Myr, $v_\mathrm{eff}=12$\,km/s. The fact that most power is concentrated at frequencies $\omega < \omega_c(k_*)$ tells us that most ISM structures are co-rotating with the local circular motion. The corresponding correlation time $\tau_* = 2\pi/\omega_c(k_*)$ associated with the wavenumber $k_*$ is thus
\begin{equation}
    \label{eq:taustar_definition}
    \tau_* \approx 16 \tau_0 \left[1 + (k_* v_\mathrm{eff}\tau_0)^2\right]^{-1/2}.
\end{equation} 
For $k_* = 0.01\,\mathrm{pc}^{-1}$, this corresponds to $\tau_* \sim 70\,$Myr.

To summarize, the picture we have in mind is of a set of ISM structures with wavelengths $\lambda_* \sim 600$ pc, roughly corotating with the guiding centers of our stars, and correlated on a timescale of $\tau_* \sim 70$ Myr. In principle, given this model for the fluctuations, we can use \autoref{eq:D_from_power} to directly evaluate the resulting diffusion coefficient. However, here we are just interested in understanding the \textit{scalings} of the diffusion coefficients with $J_R$, and so we defer such a proper mathematical derivation to C. Hamilton et al. (in preparation). Instead, we provide a heuristic argument and intuition for the scalings (though see \autoref{sec:vertical_decoupling} and \autoref{sec:environmental_dependence}, where \autoref{eq:D_from_power} will motivate a generalization of our results to galactic conditions beyond the solar neighborhood).

Consider first a very cold population, such as that corresponding to the blue curves in \autoref{fig:sigmaR_t} and \autoref{fig:rmsdJ_t} at early times in the simulation. This population will consist of stars on very \textit{small epicycles}, since their rms epicyclic amplitude $a \propto \sigma_R \propto \langle J_R \rangle^{1/2}$. As a result, the dominant potential fluctuations will satisfy $k_* a\ll 1$, which is what \cite{GK1} called the \textit{long-wavelength regime}. In this regime, stars explore only a very small part of each potential fluctuation's spatial structure over their orbits. 
The effect of the ISM fluctuations is then to impart an impulsive velocity kick $\delta \bm{v}$ to each star, independent of its epicyclic amplitude.
This gives rise to an angular momentum kick $\delta J_\varphi$ independent
of $a_R$ (see \autoref{eq:Jphi_definition}), whereas the corresponding kick in radial action is proportional to $a_R$ (see \autoref{eq:JR_definition}). 
Adding up many such random kicks, on timescales $\gg \tau_*$ we expect $D_{\varphi\varphi} \propto (\delta J_\varphi)^2$ to be independent of $a_R$ and $D_{RR} \propto (\delta J_R)^2\propto a_R^2\propto J_R$, i.e.,
\begin{equation}
    \label{eq:alpha_phi_long}
    \alpha_\varphi = 0 \quad (k_*a \ll 1)
\end{equation}
in \autoref{eq:Dphiphi_power}, and
\begin{equation}
    \label{eq:alpha_R_long}
    \alpha_R = 1 \quad (k_*a \ll 1)
\end{equation}
in \autoref{eq:DRR_power}.
Plugging $\alpha_\varphi=0$ and $\alpha_R=1$ into \autoref{eq:rms_dJR_approx} and \autoref{eq:rms_dJphi_approx}, we predict
\begin{equation}
    \label{eq:long_wavelength_scalings}
    \mathrm{rms} \,\delta J_R \propto \sigma_R^2 \propto t; \,\, \mathrm{rms} \,\delta J_\varphi \propto t^{1/2} \quad (k_*a \ll 1)
\end{equation}
for small radial actions, exactly as observed for the blue curve at early times in \autoref{fig:sigmaR_t} and \autoref{fig:rmsdJ_t}.

A more formal way to arrive at this result would have been to realize that action-space diffusion arises from stochastic ``torques'' $\dot{\bJ} \equiv \bm{\eta}(\btheta, \bJ, t)$. Formally, $\bm{\eta} = -\partial \mathcal{H}/\partial\btheta$, where $\mathcal{H}$ is the exact Hamiltonian including the ISM fluctuation potential, though at the quasilinear level $\bm{\eta}$ only needs to be evaluated along a star's unperturbed orbit ($\btheta(t) = \btheta(0)+\bm{\Omega}(\bJ)t$, $\bJ(t) = $ constant). For very cold orbits, one can then expand the angle-action space representation of $\mathcal{H}$ in a Taylor series in the small quantity $a_R/R_0$ (see, e.g., Equation B3 of \citealt{GK1}), and thereby calculate the leading-order contributions to $\eta$.  In doing so one finds $\eta_\varphi$ is independent of $a_R$ while $\eta_R\propto a_R$; squaring these and ensemble-averaging we find $D_{\varphi\varphi} \propto \eta_\varphi^2 \propto $ const and $D_{RR}\propto \eta_R^2 \propto J_R$, leading again to the scalings in \autoref{eq:long_wavelength_scalings}.

Next, as the population gets hotter, its $a$ value will increase, and it will eventually transition to the limit $k_* a\gg 1$---what \cite{GK1} termed the \textit{short-wavelength regime}. We emphasize that this transition occurs because stars evolve onto \textit{larger epicycles}; the ISM fluctuation spectrum itself remains unchanged. However, the transport behavior produced by these same fluctuations is now very different, because stars undergoing epicyclic oscillations can cross multiple peaks and troughs of the fluctuations over the course of their orbits during the timescale $\tau_*$, because $\kappa\tau_*$ is order-unity. The resulting torques will thus partially cancel each other out, so we expect transport to be suppressed, and the larger the epicycles get, the stronger this suppression should be.

Because the orbits of stars in this regime span multiple wave crests of the fluctuations, we can no longer heuristically model the diffusion as arising from uncorrelated velocity kicks which are independent of $a_R$. Instead we must turn to the more formal style of argument like that given in the paragraph below \autoref{eq:long_wavelength_scalings}, accounting for the angle-dependence of the potential fluctuations experienced by a star as it traverses its epicyclic ellipse. Following Equation B12 of \cite{GK1}, the stochastic torques in this regime scale with $a_R$ as $\bm{\eta} \propto a_R^{-1/2}$. Mathematically, this scaling follows from taking the large-argument expansion of a Bessel function, $J_{n_R}(k_* a_R) \propto (k_* a_R)^{-1/2}$ when $ka_R \gg n_R^2$. In our case, we can concentrate on $|n_R| \lesssim 2$, because for larger $n_R$, the corresponding increase in $\omega_\mathrm{res}$ (see the discussion below \autoref{eq:D_from_power}) exceeds the cutoff frequency $\sim \omega_c(k_*)$ of the ISM fluctuation power spectrum (see the discussion above \autoref{eq:taustar_definition}). Intuitively, this scaling follows from the fact that the number of wave crests of wavenumber $k_*$ that a star traverses over its epicyclic ellipse is $N \propto k_* a_R$ (provided $\tau_*\kappa \gtrsim 1$), and the number of these peaks/troughs that survive the cancellation effect is $\sim 1/\sqrt{N} \propto (k_* a_R)^{-1/2}$. Once again, the field $\bm{\eta}$ is correlated over a timescale $\tau_*$, so on timescales $\gg \tau_*$ we have $D_{\varphi\varphi} \propto \eta_\varphi^2 \propto a_R^{-1} \propto J_R^{-1/2}$, i.e.,
\begin{equation}
    \label{eq:alpha_phi_short}
    \alpha_\varphi = -\frac{1}{2} \quad (k_*a \gg 1)
\end{equation}
in \autoref{eq:Dphiphi_power}, and also $D_{\varphi\varphi} \propto \eta_R^2 \propto J_R^{-1/2}$, i.e.,
\begin{equation}
    \label{eq:alpha_R_short}
    \alpha_R = -\frac{1}{2} \quad (k_*a \gg 1)
\end{equation}
in \autoref{eq:DRR_power}. 
Plugging $\alpha_\varphi=\alpha_R=-1/2$ in \autoref{eq:rms_dJR_approx} and \autoref{eq:rms_dJphi_approx}, we predict
\begin{equation}
    \label{eq:short_wavelength_scalings}
    \mathrm{rms} \,\delta J_R \propto \sigma_R^2 \propto t^{2/5}; \,\, \mathrm{rms} \,\delta J_\varphi \propto t^{2/5} \quad (k_*a \gg 1)
\end{equation}
for larger radial actions, just as observed at late times in \autoref{fig:sigmaR_t} and \autoref{fig:rmsdJ_t}.
We also highlight the fact that the value of the heating-to-migration ratio in this regime is constant, and takes a particularly simple form: directly from \autoref{eq:rms_dJR_approx} and \autoref{eq:rms_dJphi_approx}, we have
\begin{equation}
    \label{eq:heating_migration_ratio_short}
    \frac{\mathrm{rms} \,\delta J_R}{\mathrm{rms} \,\delta J_\varphi} = \left(\frac{2d_{RR}}{d_{\varphi\varphi}}\right)^{1/2} = \left(\frac{2D_{RR}}{D_{\varphi\varphi}}\right)^{1/2}\quad (k_*a \gg 1).
\end{equation}
Thus, whether transport is ``cool'' ($\mathrm{rms} \,\delta J_R\,/\,\mathrm{rms} \,\delta J_\varphi \ll 1$) or ``hot'' ($\mathrm{rms} \,\delta J_R\,/\,\mathrm{rms} \,\delta J_\varphi  \gtrsim 1$) in this regime depends solely on the ratio of the diffusion coefficients.

We expect the transition between the long- and short-wavelength (equivalently, small- and large-epicycle) asymptotic regimes to occur when $k_*a\sim 1$, which is when $a \sim 100\,$pc which translates to $\sigma_R \sim 3$\,km/s. Inspecting \autoref{fig:sigmaR_t}, this is within a factor of 2 of the turnover value of $\sigma_R \approx 6\,$km/s for the blue ($\sigma_{R,0} = 0$) curve. Furthermore, the green ($\sigma_{R,0} = 3\,$km/s) curve in the Figure already deviates significantly from the $\propto t^{1/2}$ behavior of the blue curve, and the yellow ($\sigma_{R,0} = 5\,$km/s) curve is nearly $\propto t^{1/5}$ for the entire simulation duration, as anticipated for an ensemble too hot to ever reside in the long-wavelength regime. 

It is now also easy to understand the behavior we found in \autoref{fig:rmsdJ_t_variedA} when varying the strength of the fluctuations by a factor $A$. In particular, multiplying the fluctuations by $A$ scales the diffusion coefficients by $A^2$ without changing the value of $k_*$. The stronger (weaker) fluctuations drive the system to the same transition point of $k_*a\sim 1$ at an earlier (later) time, but the asymptotic scalings before and after this transition are the same in every case. More quantitatively, inspecting \autoref{eq:rms_dJR_approx} and \autoref{eq:rms_dJphi_approx}, we see that multiplying the diffusion coefficients by $A^2$ should shift the amount of heating $\mathrm{rms}\,\delta J_R$ by $A^2$ at early times and by $A^{4/5}$ at late times, and shift the amount of migration $\mathrm{rms}\,\delta J_\varphi$ by $A$ at early times and by $A^{4/5}$ at late times. The shifts between the magenta, blue, and cyan curves in \autoref{fig:rmsdJ_t_variedA} agree well with these predictions. Notably, similar heating-to-migration ratio values are achieved at late times across all the curves simply because the factors of $A$ cancel out in the ratio $D_{RR}/D_{\varphi\varphi}$ following \autoref{eq:heating_migration_ratio_short} for the short-wavelength regime.\footnote{The slight differences in the heating-to-migration ratio curves in \autoref{fig:rmsdJ_t_variedA} arise because the ensembles spend different amounts of time in the long-wavelength regime, where the ratio is not driven towards a constant.}

\subsection{Diffusion Coefficients for the Solar Neighborhood}
\label{sec:diffusion_measurements}

Equipped with this understanding of the power law indices $\alpha_R$ and $\alpha_\varphi$, we can now estimate the action-space diffusion coefficients for ISM-driven transport in the solar neighborhood. To do this, we measure the coefficients of the best-fit black dashed (long-wavelength) and dotted (short-wavelength) lines in panels (a) and (b) of \autoref{fig:rmsdJ_t}, holding the slopes fixed at the power law indices appropriate for each regime. Then, using \autoref{eq:rms_dJR_approx} and \autoref{eq:rms_dJphi_approx}, we relate these coefficients to the values of $d_{RR}$ and $d_{\varphi\varphi}$ in both the long- and short-wavelength regimes. In this way, we find
\begin{equation}
    \label{eq:DRR_R8}
    D_{RR}^{\text{(R8)}} \approx \begin{cases}
        9.3\,\frac{\mathrm{(kpc\,km/s)^2}}{\mathrm{Gyr}}\times\left(\frac{J_R}{\mathrm{kpc\,km/s}}\right) & (J_R \ll J_{*}) \\[4pt]
        6.9\,\frac{\mathrm{(kpc\,km/s)^2}}{\mathrm{Gyr}}\times\left(\frac{J_R}{\mathrm{kpc\,km/s}}\right)^{-\frac{1}{2}} & (J_R \gg J_{*})
    \end{cases}
\end{equation}
for the radial diffusion coefficient, and
\begin{equation}
    \label{eq:Dphiphi_R8}
    D_{\varphi\varphi}^{\text{(R8)}} \approx \begin{cases}
        6900\,\frac{\mathrm{(kpc\,km/s)^2}}{\mathrm{Gyr}} & (J_R \ll J_{*}) \\[4pt]
        4600\,\frac{\mathrm{(kpc\,km/s)^2}}{\mathrm{Gyr}}\times\left(\frac{J_R}{\mathrm{kpc\,km/s}}\right)^{-\frac{1}{2}} & (J_R \gg J_{*})
    \end{cases}
\end{equation}
for the azimuthal diffusion coefficient. We have defined
\begin{equation}
    \label{eq:Jstar_definition}
    J_* \equiv \frac{\kappa}{2k_*^2} = \frac{\kappa\lambda_*^2}{8\pi^2}
\end{equation}
to be the value of the radial action for which $k_* a_R = 1$, which is approximately $0.2\,\mathrm{kpc\,km/s}$ for the R8 simulation. As a simple consistency check for our measurements, we can plug the short wavelength regime ($J_R\gg J_*$) results from \autoref{eq:DRR_R8} and \autoref{eq:Dphiphi_R8} into \autoref{eq:heating_migration_ratio_short}, which yields the prediction $\mathrm{rms}\,\delta J_R \,/\,\mathrm{rms} \,\delta J_\varphi \approx 0.055$. This agrees nicely with the late-time heating-to-migration ratio measured in panel (c) of \autoref{fig:rmsdJ_t}.

Strictly, speaking the measured coefficients $D_{ii}^{\text{(R8)}}$ are only defined in the asymptotic regimes of very small or very large $J_R/J_*$, because the measurement relies on the assumption that the $D_{ii}$ are pure power laws of index $\alpha_i$. However, a simple prescription for the diffusion coefficients at intermediate values of $J_R \sim J_*$ could be to extrapolate them using a ``broken power-law'' model. We note that in such a model, for the diffusion coefficients to be continuous as functions of $J_R$, the transition between the asymptotic limits occurs at a value of $J_R \approx 0.82\,\mathrm{kpc}\,\mathrm{km/s} \sim 4J_*$ for $D_{RR}$, and $J_R \approx 0.45\,\mathrm{kpc}\,\mathrm{km/s} \sim 2J_*$ for $D_{\varphi\varphi}$. The fact that neither of these turnover points is precisely equal to $J_*$ is not surprising, since $k_*a\sim 1$ is only a rough estimate of the transition between asymptotic regimes, not a precise requirement. The fact that they differ \textit{between radial and azimuthal diffusion coefficients} is likely due to the fact that radial and azimuthal actions are each sensitive to different components of a spatial perturbation (C. Hamilton et al., in preparation), and the perturbations are preferentially trailing rather than isotropic (see, e.g., Figure 6 of \citealt{Modak2026characterizing}). Thus, the detailed dynamics is more complicated than we have assumed here, with our single dominant wavenumber $k_*$ and our assumption of isotropy in the fluctuations.

Subject to the above caveats, these measured diffusion coefficients may be used to approximately capture the ISM's contribution to radial transport in action space in (semi-)analytic stellar-dynamical models. To be precise, after a timestep $\Delta t$, stellar actions should be updated according to the Euler-Maruyama update rules $J_i \mapsto J_i + \Delta J_i$, where
\begin{equation}
    \label{eq:Langevin_update_step}
    \Delta J_i = X_i \left[D_{ii}(J_R) \Delta t\right]^{1/2},
\end{equation}
and $X_R$ and $X_\varphi$ are independent draws from a Gaussian distribution with zero mean and unit variance (e.g., \citealt{Spitzer1987}). Note that $\Delta t$ need not be the timestep used for integrating the unperturbed dynamics---the only requirement is that the action increments satisfy $\Delta J_i/J_i \ll 1$ at all times, so a larger value of $\Delta t$ can often be chosen as needed when implementing the diffusion for computational efficiency.

Finally, we stress that the diffusion coefficients we have measured in \autoref{eq:DRR_R8} and \autoref{eq:Dphiphi_R8} are applicable \textit{only to initially vertically cold stellar populations in present-day solar-neighborhood conditions}. We describe how the diffusion coefficients are altered by stellar vertical motions in \autoref{sec:vertical_decoupling}, and how to generalize these local results to other galactic environments and connect them to the global context in \autoref{sec:environmental_dependence}.

\subsection{Modeling ISM Transport using Isotropic GRFs}
\label{sec:GRF_model}

The transport results uncovered in \autoref{sec:results} are very different from classical models that treat the ISM's dynamical influence as arising from compact, spherically-symmetric ``GMCs" treated as, e.g., Plummer spheres (see also \autoref{sec:comparison_conventional}). A natural question is then, can we ``mock-up'' the more realistic fluctuations present in the TIGRESS-NCR simulations with a simple semi-analytic prescription? Ideally, such mock ISM fluctuations would retain the convenience of spherical cloud 
models (i.e., they would not require a full treatment of gas physics) while still producing correct transport results.

One simple prescription for generating more realistic mock ISM fluctuations is to create isotropic Gaussian random field (GRF) realizations of the ISM density field that approximately match the measured one- and two-point statistics of the gas density in the TIGRESS-NCR simulations. \cite{Modak2026characterizing} provide analytic models for the surface density spatio-temporal power spectra of both the R8 and LGR4 simulations (see their Equation 32), and subsequently connect the surface density spectra to power spectra of the gravitational potential by specifying a model for the vertical structure of the gas (see their Equation 54). These models are idealized in that they do not account for (i) the slight anisotropic preference for trailing structures in the ISM at large scales, and (ii) the non-Gaussianity of ISM structures, which do not have random phases because they are nonlinear in amplitude and at small scales are filamentary. For the purposes of this work, we (i) adopt the isotropic model for the ISM power spectrum simply for its convenience (and will demonstrate its adequacy in modeling radial transport below), and (ii) note that in quasilinear diffusion theory (which is an excellent approximation in this context), following \autoref{eq:D_from_power}, only the fluctuation power spectrum (and not phase information) enters the diffusion coefficient calculation, so for the purposes of getting transport right, we are free to neglect non-Gaussian features of the ISM.

To test whether fluctuation fields generated using these \textit{spatio-temporal} model spectra can match the transport behavior induced by the true TIGRESS-NCR fluctuations, we follow Appendix B of \cite{Modak2026characterizing} to produce a mock realization\footnote{To be precise, we generate realizations following the ``exponential time-dependence'' method described in Appendix B.2 of \cite{Modak2026characterizing} with $N_p = 2t_\mathrm{sim}/\tau_0$ and $t_\mathrm{buffer} = 2\tau_0$; we find no difference in transport results for ensembles integrated in fluctuations produced using the more standard GRF realization method described in Appendix B.1.} of solar-neighborhood ISM surface density fluctuations (i.e., using the specified values for the R8 simulation parameters from their Table 1 through Table 5), and translate them to gravitational potential fluctuations using their fitted vertical profiles (see their Equation 48).\footnote{We assume sinusoidal time-dependence of the vertical center of mass position $z_\mathrm{com}$ and scale height $H$ following the measured amplitudes in their Table 4, but have checked that the precise time-dependence of the ISM's vertical structure does not materially alter the radial transport. Of course, the ISM's vertical structure \textit{does} play a significant role in determining the \textit{vertical} transport it drives; we will analyze this in Paper II.} Additionally, in order to test the importance of the finite correlation time of ISM structures, we generated mock fluctuations with the same \textit{spatial} model spectra as measured in \cite{Modak2026characterizing}, but applied a new realization every $1\,$Myr, effectively giving the ISM fluctuations a much shorter correlation time. These white-noise-like fluctuation fields are much more computationally straightforward to generate than those with nontrivial temporal structure, so it is interesting to know whether the transport they drive is significantly different.

\begin{figure}
    \centering
    \includegraphics[width=0.47\textwidth]{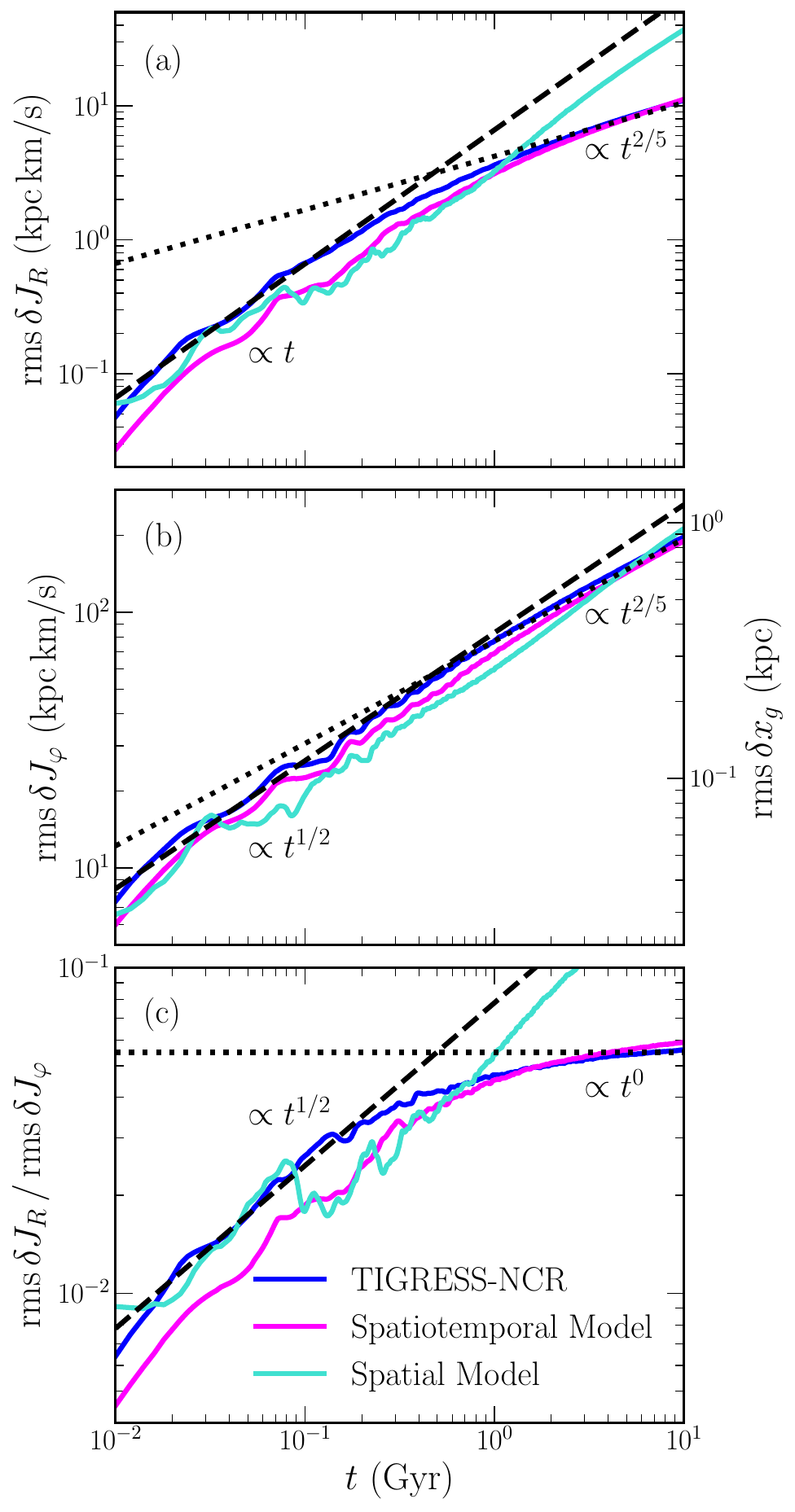}
    \caption{As in \autoref{fig:rmsdJ_t}, except only for initially cold ($\sigma_{R,0} = 0$) ensembles, integrated in ISM fluctuations with the model spatio-temporal power spectrum from \cite{Modak2026characterizing} (magenta), and in ISM fluctuations with the model spatial spectrum from \cite{Modak2026characterizing}, but with white-noise temporal structure (cyan). For comparison, we also show the ensemble integrated in the real TIGRESS-NCR R8 fluctuations (blue, same as the blue curves in \autoref{fig:rmsdJ_t}).}
    \label{fig:rmsdJ_t_GRFcomparison}
\end{figure}

In \autoref{fig:rmsdJ_t_GRFcomparison}, we show the results of integrating ensembles of test particles through these mock fluctuation fields using the same method and $\sigma_{R,0} = 0$ initial conditions as the fiducial cold simulation presented in \autoref{sec:results} (once again shown in blue for comparison). In magenta, we show the results for the ensemble experiencing the full spatio-temporally-correlated model fluctuations, while in cyan we show the results for the case of spatially correlated but temporally uncorrelated fluctuations.

We see that the magenta (spatio-temporal model) curves exhibit many of the same features of the blue (TIGRESS-NCR potential) curves, including a turnover from $\mathrm{rms}\,\delta J_R \propto t$ to $\propto t^{2/5}$ and from $\mathrm{rms}\,\delta J_\varphi \propto t^{1/2}$ to $\propto t^{2/5}$, and thus the transport remains ``cool,'' i.e., the heating-to-migration ratio at late times is $\ll 1$. The turnovers also occur at approximately the same point in both the magenta and blue curves, indicating that the model spatial fluctuation statistics are adequately capturing the importance of the characteristic scale $k_*$ in driving transport.

However, the cyan curves (arising from temporally-uncorrelated fluctuations drawn from the specified spatial power spectrum) differ substantially. They always exhibit the scalings of the long-wavelength regime, and never transition to the more gradual scalings of the short-wavelength regime, even though they share the same spatial spectrum (and thus the same dominant spatial scale $\lambda_*$). This is simple to understand physically: because the fluctuations de-correlate on a very short timescale, the stars have no chance to experience multiple wave crests of a given fluctuation field before it is replaced by a new realization. Thus, the cancellation of torques that drove the more gradual transport characteristic of the short-wavelength regime (see \autoref{sec:physical_interpretation}) can never occur in these models. Quantitatively, this manifests in significantly increased heating after $10$\,Gyr, and a heating-to-migration ratio that grows roughly as $\propto t^{1/2}$ for all times. These erroneous results demonstrate the importance of properly accounting for not only the spatial, but also the temporal, structure of the ISM in dynamical modeling.

\section{Discussion}
\label{sec:discussion}

In \autoref{sec:solar_neighborhood_sims}, we presented the outcomes of orbit integrations through our fiducial solar-neighborhood ISM fluctuations, used quasilinear diffusion theory to interpret the resulting radial transport, and demonstrated that realistic ISM-driven transport can be (approximately) reproduced with mock fluctuation fields that are calibrated to (approximately) match the ISM's two-point spatio-temporal statistics. Below, we quantify the effects of stellar vertical motions on radial transport (\autoref{sec:vertical_decoupling}), compare our results with those of more conventional ``compact cloud'' models for the ISM's dynamical influence (\autoref{sec:comparison_conventional}), discuss how to scale our local solar-neighborhood results to other Galactic contexts (\autoref{sec:environmental_dependence}), and assess the ISM's role in contributing to the radial heating and migration observed in the Galaxy (\autoref{sec:comparison_observations}).

\subsection{Effects of Vertical Motions on Radial Transport}
\label{sec:vertical_decoupling}

\begin{figure}
    \centering
    \includegraphics[width=0.47\textwidth]{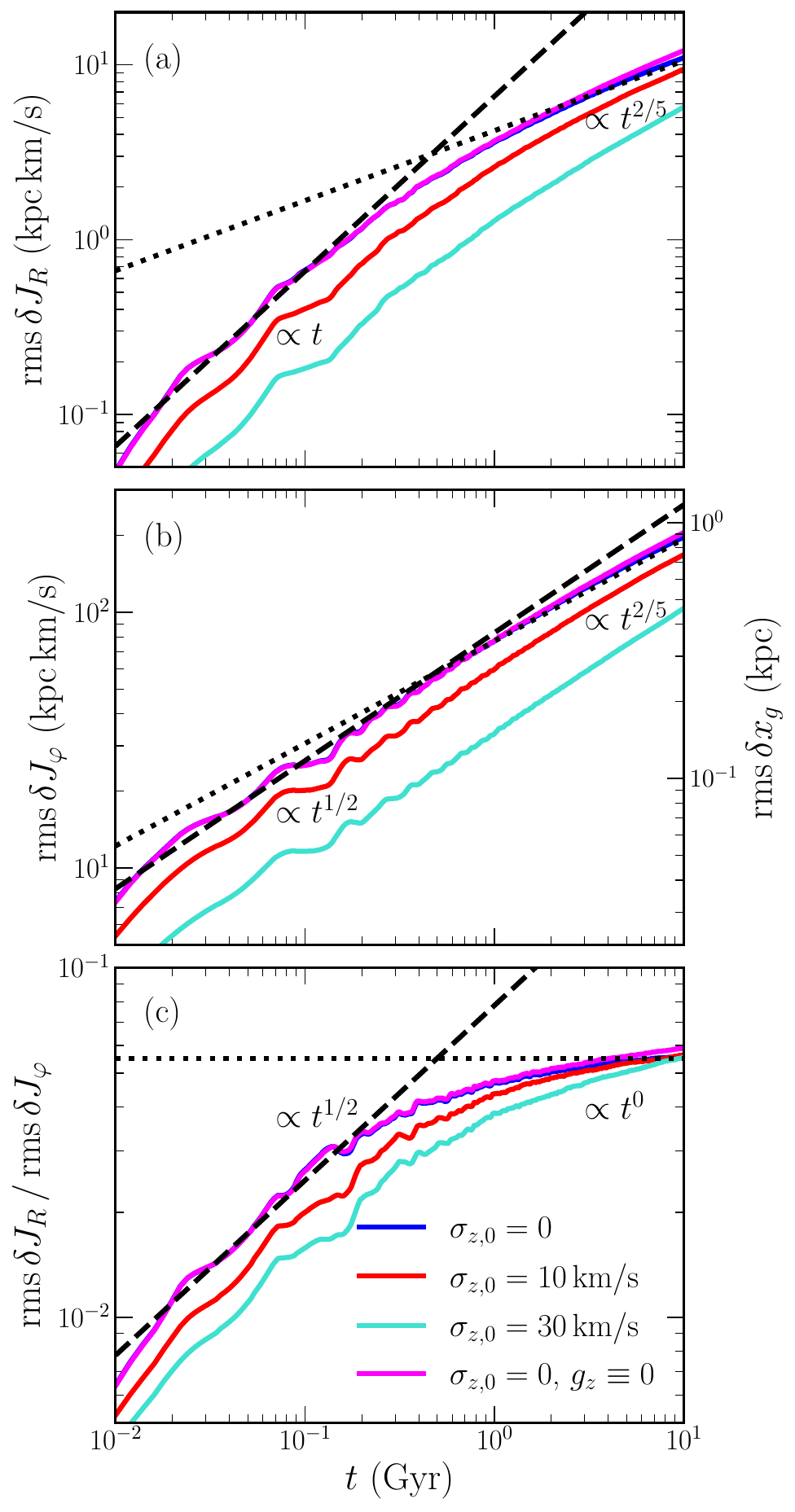}
    \caption{As in \autoref{fig:rmsdJ_t}, except only for initially radially cold ($\sigma_{R,0} = 0$) ensembles with varied vertical initial conditions: $\sigma_{z,0} = 0$ (blue, same as the blue curves in \autoref{fig:rmsdJ_t}, almost entirely hidden by the magenta curves), $10\,$km/s (red), and $30\,$km/s (cyan). For comparison, we also show results for an ensemble with $\sigma_{R,0} = \sigma_{z,0} = 0$ with vertical forcing turned off entirely, so that $\sigma_z = 0$ at all times (magenta).}
    \label{fig:rmsdJ_t_verticaldecoupled}
\end{figure}

Throughout \autoref{sec:solar_neighborhood_sims}, we focused on radial heating and migration in initially vertically cold ensembles (i.e., stars with $\sigma_{z,0} = 0$). Here, we illustrate how these radial transport processes are altered by the inclusion of larger vertical stellar motions. In \autoref{fig:rmsdJ_t_verticaldecoupled}, we compare the rms radial and azimuthal action changes produced in our fiducial $\sigma_{R,0} = \sigma_{z,0} = 0$ simulation (blue) to results from vertically warmer simulations with $\sigma_{R,0} = 0$, but $\sigma_{z,0} = $10\,km/s (red) and 30\,km/s (cyan). We also show results for a simulation with $\sigma_{R,0} = \sigma_{z,0} = 0$, but with vertical accelerations from the ISM turned off entirely, which we denote ``$g_z \equiv 0$'' (magenta). In this case, stars are confined to the midplane with $\sigma_z = 0$ throughout.

We see that all the curves share essentially the same power-law asymptotic scalings, but with reductions in overall amplitude of the transport as the stellar vertical motions increase in amplitude (red compared to blue, and cyan compared to red). Additionally, the vertical and radial transport driven by the ISM are well-approximated as \textit{decoupled}: including the vertical forcing from the ISM in our fiducial ensembles only leads to minor differences from the ensemble with vertical forcing switched off (blue compared to magenta).

The reduction in overall amplitude of both the heating and migration for vertically warmer disks has a straightforward, intuitive interpretation: as the vertical velocity dispersion increases, so does the typical vertical epicyclic amplitude of the ensemble (see \autoref{eq:rms_vertical_velocity}). Stars in vertically warmer ensembles spend a smaller fraction of their orbital periods near the midplane, where the ISM potential fluctuations are strongest, and thus experience reduced scattering. We defer a discussion of the amplitude and scaling of ISM-driven vertical heating to Paper II, while below we describe a general model for the role of vertical motions in suppressing radial transport.

Quantitatively, the effective diffusion tensor governing radial transport in \autoref{eq:quasilinear_diffusion} can be replaced by its average value over a typical star's vertical orbit, which will be reduced compared to its value for orbits confined to the midplane by a (scalar) ``suppression factor'' $\xi$:
\begin{equation}
    \label{eq:suppression_definition}
    D_{ij}(J_\varphi, J_R, J_z)  \simeq \xi(J_z) D_{ij}(J_\varphi, J_R, 0).
\end{equation}
Since the diffusion tensor is proportional to the power spectrum of potential fluctuations (see \autoref{eq:D_from_power}), the problem of calculating $\xi$ can be reduced to analyzing the dependence of the spectrum $P_{\delta\phi}$ on $z$, which has been measured for the TIGRESS-NCR simulations we use here by \cite{Modak2026characterizing}. They demonstrate that if the gas volume density fluctuations take the form $\delta\rho(x, y, z) = \delta\Sigma(x, y)\zeta(\ztilde)/(2h)$ for a dimensionless vertical profile $\zeta$, following their Equation 48, the $z$-dependence of the spectrum is $P_{\delta\phi} \propto \chi(\ktilde, \ztilde)^2$, where
\begin{equation}
    \label{eq:chi_definition}
    \chi(\ktilde, \ztilde) \equiv \frac{1}{2} \int \md \ztilde' \, e^{-\ktilde|\ztilde - \ztilde'|}\zeta(\ztilde').
\end{equation}
In these expressions, $\ztilde \equiv (z-z_\mathrm{com})/H$ is the distance to the vertical center of mass position of the gas layer $z_\mathrm{com}$ in units of its rms thickness $H$, $\ktilde \equiv k H$, and the factors of $2h \equiv \int \md z \,\zeta$ ensure that the integral of $\delta\rho$ over $z$ yields $\delta\Sigma$, as required by the definition of surface density. Thus, performing the vertical orbit average using the epicyclic approximation of \autoref{eq:shearing_trajectory_z}, we can calculate the suppression factor as
\begin{align}
    \label{eq:xi_definition}
    \xi(J_z, k_*, H) & \simeq \frac{1}{\chi(\ktilde_*, 0)^2}\int \frac{\md\theta_z}{2\pi} \, \chi \! \left(\! \ktilde_*, -\sqrt{\frac{2J_z}{\nu H^2}}\cos\theta_z \! \right)^2 \! \!,
\end{align}
where $\ktilde_* \equiv k_* H$. Although we have written the arguments of $\xi$ out explicitly for clarity, it only depends on the combinations $k_* H$ and $[J_z/(\nu H^2)]^{1/2}$. Note also that $\xi = 1$ whenever $J_z = 0$, as it must given the definition in \autoref{eq:suppression_definition}.

\begin{figure}
    \centering
    \includegraphics[width=0.47\textwidth]{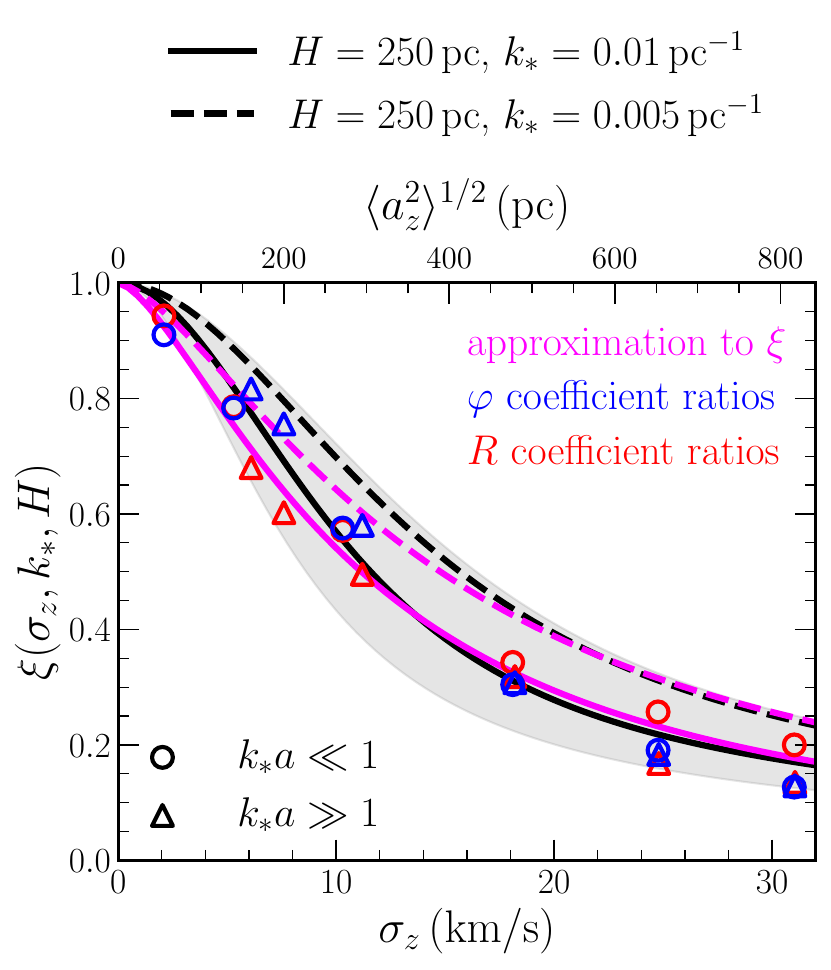}
    \caption{The vertical suppression factor $\xi$ of the diffusion tensor in \autoref{eq:suppression_definition} as a function of the stellar population's vertical velocity dispersion (lower axis) or vertical epicyclic amplitude (upper axis), calculated for ISM fluctuations with a vertical profile given by \autoref{eq:zeta_definition} using the R8 parameter values. The solid black curve is for our fiducial value of $k_* = 0.01\,\mathrm{pc}^{-1}$ with $H = 250\,$pc, while the shaded grey region spans $H = 150\,\mathrm{pc}$ to $420\,\mathrm{pc}$ for the same $k_*$. We also show the suppression for $k_* = 0.005\,\mathrm{pc}^{-1}$ and $H = 250\,$pc as a black dashed curve. The analytic approximation to $\xi$ given in \autoref{eq:xi_approx} is shown in magenta for both solid and dashed curves. Measured ratios of diffusion coefficients for simulations with varied $\sigma_{z,0}$ relative to the $g_z \equiv 0$ simulation are overplotted, with azimuthal (radial) coefficient ratios shown in blue (red). The circle/triangle symbols correspond to long-/short-wavelength regime measurements, i.e., measurements at early/late times.}
    \label{fig:vertical_suppression}
\end{figure}

The precise form of $\xi$ depends on the choice of vertical density profile $\zeta$. Figure 11 of \cite{Modak2026characterizing} demonstrates that the vertical density profile in the TIGRESS-NCR R8 simulation is well-fit by a weighted mixture of $\sech^2$ (the equilibrium for an isothermal self-gravitating layer) and exponential (the equilibrium for an isothermal layer confined in an external field) profiles:
\begin{equation}
    \label{eq:zeta_definition}
    \zeta(\ztilde) = (1-w)\sech^2\left(\frac{\ztilde}{\alpha_\mathrm{s}}\right) + we^{-|\ztilde|/\alpha_\mathrm{e}},
\end{equation}
with $w = 0.38$, $\alpha_\mathrm{s} = 0.64$, and $\alpha_\mathrm{e} = 0.82$. For reference, for this profile, $h = \alpha_w H$, where $\alpha_w \equiv (1-w)\alpha_s + w\alpha_e \approx 0.71$. Unfortunately, for this choice of $\zeta$ (and most other physically motivated profile choices), no simple analytic expression exists for the integral in \autoref{eq:xi_definition}, so it must be computed numerically in general. However, we find that the fitting formula
\begin{equation}
    \label{eq:xi_approx}
    \xi \approx \frac{(1 + \alpha_w k_* H)^2}{1 + \left[(1 + \alpha_w k_* H)^2 - 1\right]\left[1 + (k_* H)^{\frac{1}{3}} \left(\frac{2J_z}{\nu H^2}\right)^{\frac{3}{4}}\right]},
\end{equation}
retains the correct limiting behavior in both $k_*$ and $J_z$, with an absolute error of $ \lesssim 0.07$ (a relative error of $\lesssim 8\%$) across the relevant parameter space for the R8 vertical profile---see the magenta curves in \autoref{fig:vertical_suppression}.

In \autoref{fig:vertical_suppression}, we plot the suppression factor $\xi$ given by \autoref{eq:xi_definition} with $k_* = 0.01\,\mathrm{pc}^{-1}$ as a function of $\sigma_z$ ($=(\nu \langle J_z \rangle)^{1/2}$, see \autoref{eq:rms_vertical_velocity}) for $H = 250\,$pc (solid black curve) and $H = 150-420\,$pc (shaded grey region). These $H$ values correspond to approximately the time-averaged, minimum, and maximum\footnote{For simplicity, we have fixed $z_\mathrm{com} = 0$ in the Figure as well as the analysis leading to \autoref{eq:xi_definition}, but have roughly accounted for its contribution by extending the range of $H$ values by its rms variation of $\sim 20\,$pc; see, e.g., Figure 10 and Table 4 of \cite{Modak2026characterizing}.} values measured in \cite{Modak2026characterizing}. We also plot ratios of the radial and azimuthal diffusion coefficients (triangles and circles respectively) measured in the long- and short-wavelength regimes (filled and unfilled symbols respectively) for simulations with varied $\sigma_{z,0}$, relative to the diffusion coefficients measured in the simulation with vertical forcing turned off entirely, so that $J_z = 0$ at all times (see \autoref{eq:suppression_definition}). To determine the value of $\sigma_z$ to plot on the horizontal axis for each simulation, we calculate the time-averaged value of $\sigma_z$ over the duration in which the ensemble is in each wavelength regime; the simulations with $\sigma_{z,0} \gtrsim 10\,$km/s remain essentially fixed at their initial $\sigma_z$ values throughout the simulation.\footnote{The $\sigma_{z,0} = 20\,$km/s, $\sigma_{z,0} = 30\,$km/s, and $\sigma_{z,0} = 40\,$km/s simulations actually have initial vertical velocity dispersions of $\approx 18\,$km/s, $\approx 24\,$km/s, and $\approx 31\,$km/s respectively---for such large values of $\sigma_{z,0}$, the vertical epicyclic approximation we use when sampling initial conditions as in \autoref{eq:schwarzschild_IC_z} breaks down, so the actual value of $\sigma_z(t = 0)$ is slightly less than the parameter $\sigma_{z,0}$.}

We see that while the measured diffusion coefficient ratios do exhibit some scatter, they are largely consistent with the suppression predicted using \autoref{eq:xi_definition} for the range of $H$ values present in the TIGRESS-NCR simulation. Additionally, as expected, the measured ratios are similar across both long- and short-wavelength regimes, and for both radial and azimuthal diffusion coefficients. Thus, to model radial transport in stellar populations with warmer initial conditions than those we have studied in \autoref{sec:results}, we may still use the diffusion coefficients as measured in e.g., \autoref{eq:DRR_R8} and \autoref{eq:Dphiphi_R8} for $\sigma_{z,0} = 0$ simulations, except they should be multiplied by an additional suppression factor $\xi$ according to their typical vertical velocity dispersion. The formula given in \autoref{eq:xi_approx} provides a good approximation for the suppression factor needed in a present-day, solar-neighborhood-like ISM. In \autoref{sec:environmental_dependence}, we discuss further how to generalize these results to ISM conditions in other galactic environments.

\subsection{Comparison with Conventional ISM Models}
\label{sec:comparison_conventional}

The radial heating and migration we measured in \autoref{sec:results} differs both qualitatively and quantitatively from the transport behavior induced by conventional ``GMC'' models adopted in the dynamics community. Inheriting the conceptual framework introduced by \cite{SpitzerSchwarzschild1951, SpitzerSchwarzschild1953}, the ISM's contribution to heating and migration processes in the Galaxy has typically been modeled as a series of impulsive kicks with compact ($\sim 20\,$pc), spherically-symmetric perturbers (e.g., Plummer spheres). We will now briefly review the main features of the conventional models, in order to contrast them with the interpretation we presented in \autoref{sec:physical_interpretation} (for further details, see Section 8.4.1 of \citealt{BT} and references therein).

Consider a star with an epicyclic amplitude $a_R$ encountering a cloud of mass $m_\mathrm{c}$ at an impact parameter $b$ with relative velocity $v$. The impulse approximation tells us that the star will receive an absolute velocity kick of $\sim Gm_\mathrm{c}/(bv)$. If there are $n$ such clouds per unit volume, then a standard ensemble- and encounter-averaged calculation yields a heating rate
\begin{equation}
    \label{eq:heating_clouds_estimate}
    \frac{\md \sigma_R^2}{\md t} \sim \left\langle \int 2\pi b \,\md b\,  nv\,\left( \frac{Gm_\mathrm{c}}{bv} \right)^2 \right\rangle.
\end{equation}
First, if $a_R \ll b$ (i.e., the star is on a very cold orbit or the encounter is very distant), the encounter speed $v$ will be dominated by the \textit{shear}: $v \sim q\Omega b$, independent of $a_R$ (and hence of $\sigma_R$ after ensemble-averaging over many such stars). Then the right hand side of \autoref{eq:heating_clouds_estimate} is independent of $\sigma_R$, so we find $\sigma_R\propto t^{1/2}$. However, as the star's orbit becomes warmer, the encounter speed $v$ will come to be dominated by the radial motions, i.e., the \textit{dispersions}: $v \sim \kappa a_R$. Then the right hand side of \autoref{eq:heating_clouds_estimate} is proportional to $\sigma_R^{-1}$ after ensemble averaging, implying $\sigma_R\propto t^{1/3}$. The transition between the shear- and dispersion-dominated regimes occurs when $\sigma_R \sim q\Omega r_J \sim 1.5\,\mathrm{km/s} \times (m_\mathrm{c}/(10^5\,\Msun))^{1/3}$ for the solar neighborhood, where $r_J \equiv (Gm_\mathrm{c}/(2q\Omega))^{1/3}$ is the cloud's Jacobi radius. Thus, according to these calculations, real stellar populations would almost always be in the dispersion-dominated regime and would be expected to exhibit a $\sigma_R\propto t^{1/3}$ scaling across essentially all times, in contrast with our findings of $\sigma_R\propto t^{1/5}$ at late times.

We note that this heuristic calculation assumes encounters are confined to the plane and neglects the effect of vertical scattering on the planar dynamics. By incorporating contributions from the vertical velocity dispersion $\sigma_z$ into the encounter speed and solving coupled equations for the radial and vertical heating, works by \cite{Lacey1984, Ida1990, KokuboIda1992} found a reduced scaling of $\sigma_R \propto t^{1/4}$, demonstrating that not only the amplitude, but also the scaling of cloud-driven radial transport is altered by stellar vertical motions.\footnote{Their cloud-driven transport model requires coupling the vertical and radial heating rates, so this reduced scaling is only self-consistent with a vertical velocity dispersion scaling as $\sigma_z \propto t^{1/4}$ as well. We will explicitly compare this predicted vertical heating rate to that of our simulations in Paper II.} Again, this contrasts with our finding that the radial transport scalings remain essentially unchanged by stellar vertical motions, and only the amplitude differs.

Furthermore, in the ``compact perturber'' model, transport tends to be ``hot'' in both the shear- and dispersion-dominated regimes, in the sense that the heating-to-migration ratio tends towards large values. For instance, if we encapsulate the encounter microphysics in a mean-squared velocity kick $\langle (\delta \bvee)^2 \rangle$, the amount of migration driven is (see \autoref{eq:Jphi_definition}) $\mathrm{rms}\,\delta J_\varphi \sim R_0 \langle (\delta \bvee)^2 \rangle^{1/2}$, while the amount of heating is (see \autoref{eq:JR_definition}) $\mathrm{rms}\,\delta J_R \sim \langle (\delta \bvee)^2 \rangle/\kappa$. Therefore, the heating-to-migration ratio continuously grows as stars migrate: $\mathrm{rms}\,\delta J_R\,/\,\mathrm{rms}\,\delta J_\varphi \propto \mathrm{rms}\,\delta J_\varphi$. Note that this contrasts starkly with our finding that $\mathrm{rms}\,\delta J_R\,/\,\mathrm{rms}\,\delta J_\varphi$ approaches a small constant value ($\approx 0.055$) at late times: real ISM-driven transport remains ``cool.''

Owing to estimates like those discussed above, which assume that GMCs are the dominant contributors to ISM-driven heating and migration, the ISM was deemed an unlikely driver of the substantial radial transport observed in the Galaxy (see \autoref{sec:comparison_observations}), which was instead attributed primarily to large-scale spiral arms (e.g., \citealt{BinneyLacey1988}). Instead, the chief role assigned to the ISM under these traditional compact cloud models was to shape the velocity ellipsoid by redirecting in-plane energy (produced by the spiral arms) into vertical motion \citep{Lacey1984, Carlberg1987, JenkinsBinney1990, Jenkins1992, BinneySellwood2002}. We will assess this notion directly in Paper II.

\begin{figure}
    \centering
    \includegraphics[width=0.47\textwidth]{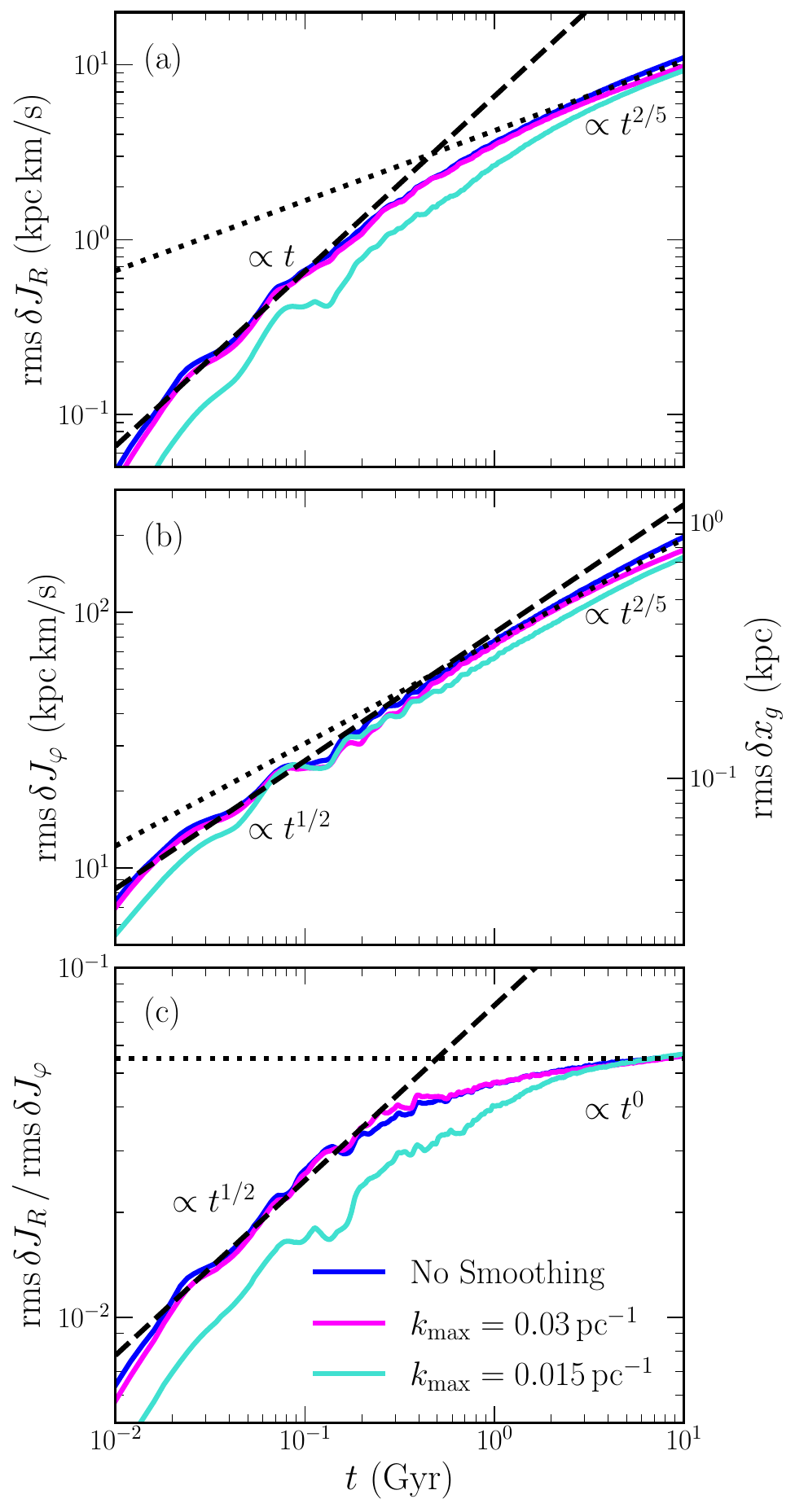}
    \caption{As in \autoref{fig:rmsdJ_t}, except only for initially cold ($\sigma_{R,0} = \sigma_{z,0} = 0$) ensembles integrated in ISM fluctuation fields smoothed using a sharp $k$-space filter with $k_\mathrm{max} = 0.03\,\mathrm{pc}^{-1}$ (magenta) and $0.015\,\mathrm{pc}^{-1}$ (cyan). For comparison, we also show our fiducial results with no smoothing applied (blue, same as the blue curves in \autoref{fig:rmsdJ_t}).}
    \label{fig:rmsdJ_t_smoothing}
\end{figure}

We have argued in this paper that one needs to model gaseous structures in the ISM in a more realistic way to get the transport behavior correct. Realistic ISM structures at small scales take the form of filaments (e.g., \citealt{Hacar2023} and references therein), and at large scales are likely better modeled as extended ($\lambda_* \sim 600\,$pc), coherent shearing wavelets (e.g., \citealt{JulianToomre1966, KimOstriker2001}) rather than compact clouds. Of course, compact cloud-like structures \textit{do} develop at small scales in the TIGRESS-NCR simulations we use in this work, and this is where star formation occurs. While present, these clouds contribute very little to the total fluctuation power---see e.g., Figure 7 of \cite{Modak2026characterizing}. They also contribute negligibly to the orbital transport. To demonstrate this explicitly, in \autoref{fig:rmsdJ_t_smoothing} we show the heating and migration produced in initially cold ($\sigma_{R,0} = \sigma_{z,0} = 0$) ensembles integrated through smoothed versions of the TIGRESS-NCR R8 fluctuation fields. These smoothed fields are generated by applying sharp (planar) $k$-space filters to the gravitational potential fluctuations $\delta\phi$, i.e., setting $\delta\phi_{\bk} = 0$ if $k > k_\mathrm{max}$, which corresponds approximately to smoothing over scales smaller than $\lambda_\mathrm{min} \equiv 2\pi/k_\mathrm{max}$. We see that setting $k_\mathrm{max} = 0.03\,\mathrm{pc}^{-1}$ ($\lambda_\mathrm{min} \sim 200\,$pc, magenta) leads to a negligible difference compared to the unsmoothed ensemble (identical to the blue curves in \autoref{fig:rmsdJ_t}), even though all compact cloud structures have been wiped out by the smoothing process. Indeed, substantial differences are only evident after smoothing over much larger scales of $k_\mathrm{max} = 0.015\,\mathrm{pc}^{-1}$ ($\lambda_\mathrm{min} \sim 400\,$pc, cyan), and even in this case, the same asymptotic behavior predicted by our interpretation in \autoref{sec:physical_interpretation} is approximately obtained.

We conclude that, rather than scattering off compact clouds, the dominant transport mechanism in our simulations is far more similar to that driven by spiral waves: stars approximately co-rotate with the local ISM structures which perturb their orbits most strongly.\footnote{In fact, the scalings of the radial heating rates $\sigma_R\propto t^{1/2}$ (long-wavelength) and $\sigma_R\propto t^{1/5}$ (short-wavelength) were already found in the specific context of heating by transient spiral waves by \cite{CarlbergSellwood1985}; here we have demonstrated that they hold much more generally and may be applied to ISM perturbations as well.} Thus, especially in the short wavelength regime, perturbations are best modeled as coherent rather than impulsive, and stars are able to experience multiple peaks and troughs of a given fluctuation, leading to an approximate cancellation of forces (see \autoref{sec:physical_interpretation}). This process, analogous to migration driven by corotation resonances with spiral perturbations, also intuitively explains the substantially ``cooler'' migration produced by a realistic ISM relative to compact-cloud models. In many circumstances, ISM-driven transport can even be \textit{cooler} than that driven by transient spiral waves, because (i) the volume-filling nature of the gas fluctuations means that stars are more frequently perturbed than in, e.g., a few-armed grand design spiral wave, and (ii) the smaller amplitude of the perturbation together with its shorter correlation time means that effects like resonance overlap (which drives additional heating; see, e.g., \citealt{Daniel2019}) do not occur. Finally, the fact that vertical structure only acts to suppress in-plane transport without altering its scaling may intuitively be understood by the fact that the resonant interactions only depend on the radial epicyclic motion, and vertical heating only serves to alter the time a star spends in the plane (see \autoref{sec:vertical_decoupling}) rather than contributing explicitly to the encounter velocity for an impulsive kick.

Ultimately, the unique radial heating and migration behavior produced by a realistic ISM suggests that other stellar-dynamical calculations grounded in conventional cloud-based models for ISM structure should be revisited. Recent simulations incorporating gas using hydrodynamics (albeit, neglecting important aspects of ISM physics like magnetic fields and accurate, self-consistent feedback, which may result in a different spatio-temporal fluctuation spectrum) in addition to stars have already found an increase in migration efficiency compared to gas-free simulations \citep{Fujimoto2023, Zhang2025}. The inclusion of gas has also been demonstrated to significantly impact the formation and evolution of the Galaxy's thick disk \citep{BlandHawthorn2025} and phase spiral \citep{TepperGarcia2025}. The theoretical model for the ISM's dynamical influence presented in this section and in \autoref{sec:physical_interpretation} above should aid in the interpretation of these results.

\subsection{Environmental Dependence of ISM-driven Transport}
\label{sec:environmental_dependence}

In \autoref{sec:solar_neighborhood_sims}, we have utilized \textit{local} shearing box models to study ISM-driven orbital transport \textit{in the solar neighborhood}; however, the radial heating and migration of stars across a galactic disk is ultimately a \textit{global} problem. There are two main reasons for this. First, in galaxies like the Milky Way, there often exist potential fluctuations on scales much larger than a kiloparsec (e.g., stellar spiral arms) and these too drive substantial transport. Second, we know that individual stars may migrate several kiloparsecs in guiding radius during their lifetimes, experiencing widely varying galactic environments, whereas our TIGRESS-NCR R8 models are designed to mimic present-day solar-neighborhood conditions only. Thus, if we are going to properly interpret observational data (see \autoref{sec:comparison_observations}) and/or provide reliable sub-grid prescriptions for ISM-driven transport in numerical galaxy modeling, we must understand how our results generalize to galactic environments beyond the solar neighborhood, in order to embed our local results into the global context. This is a challenging problem whose full solution is beyond the scope of this paper. However, we can already make several useful statements.

As discussed in \autoref{sec:comparison_conventional}, our results demonstrate that classical global models treating the ISM as a collection of compact, idealized ``clouds'' are highly unrealistic. Instead, if one is going to add ISM-driven transport to a global model, one may either proceed (i) as in \autoref{sec:GRF_model}, by generating potential fluctuations with the correct spatio-temporal power spectrum, or (ii) as described in \autoref{sec:diffusion_measurements}, by adding a stochastic Langevin forcing consistent with the desired diffusion coefficients to the equations of motion for each star (see \autoref{eq:Langevin_update_step}). The simplest approach to generalizing our results to other environments or global studies is to assess how (i) the spatio-temporal spectrum and hence (ii) the diffusion coefficients vary with the fundamental parameters that must be rescaled to suit different environments. Specifically, below we describe how these quantities scale with the mean gas surface density $\overline{\Sigma}$, the characteristic fluctuation wavelength $\lambda_*$, the fluctuation correlation time $\tau_*$, the galactocentric radius $R$, the orbital frequencies $\Omega$ and $\kappa$, and the parameters governing the vertical structure of the gas.

First, from Equation 32 and Equation 54 of \cite{Modak2026characterizing}, the ISM gravitational potential fluctuation spectrum scales as
\begin{align}
    \label{eq:P_scalings}
    P_{\delta\phi}(k, \omega) & \propto \left(\frac{G\overline{\Sigma}}{\alpha_\mathrm{w}}\right)^2 \frac{P_\delta(k, \omega) \chi^2}{k^2} \nn \\
    & \propto (G\overline{\Sigma}\lambda_*)^2 \tau_* \times \lambda_*^{n_\delta} \frac{\chi^2}{\alpha_\mathrm{w}^2},
\end{align}
where $P_\delta(k, \omega)$ is the spatio-temporal spectrum of the linear surface density fluctuation $\delta \equiv (\Sigma - \overline{\Sigma})/\overline{\Sigma}$, $n_\delta = 2.3$ is the power-law index of the spatial component of the spectrum measured from simulations (see \autoref{eq:Pdelta_scaling}), $\chi$ is the function encapsulating the ISM's vertical structure defined in \autoref{eq:chi_definition}, and $\alpha_\mathrm{w} \equiv h/H$ relates the scale height $h$ of the gas layer to its rms thickness $H$ (see the discussion surrounding \autoref{eq:zeta_definition}). In the second line, we have assumed that the dominant fluctuations are those with typical wavelength $\lambda_*$ and correlation time $\tau_*$. We have also explicitly separated out the term $(G\overline{\Sigma}\lambda_*)^2\tau_*$ for reference, as it is the ``natural unit'' for an action-space diffusion coefficient (i.e., its dimensions are (action)$^2$/time) for fluctuations with scale $\lambda_*$ and correlation time $\tau_*$.

Next, we may use \autoref{eq:D_from_power} to translate these scalings to the diffusion coefficients, calibrating against the values measured from our fiducial simulations in \autoref{eq:DRR_R8} and \autoref{eq:Dphiphi_R8}. To begin, we note that the approximate transition radial action value $J_*$ separating the long- and short-wavelength asymptotic regimes\footnote{We also recall from \autoref{sec:diffusion_measurements} that the precise turnover action values we measured for a ``broken power law'' approximation for the diffusion coefficients were not exactly at $J_R=J_*$ but instead at values of $J_R/J_* \sim 2-4$, and we expect this to remain true in different environments also.} scales as
\begin{equation}
    \label{eq:Jstar_calibrated}
    J_* \approx 0.2\,\mathrm{kpc\,km/s} \times \left(\frac{\lambda_*}{600\,\mathrm{pc}}\right)^{2} \left(\frac{\kappa}{40\,\mathrm{km/s/kpc}}\right).
\end{equation}

In the long-wavelength regime $J_R\ll J_*$, we calculate $\tilde{P}_{\delta\phi}$ by squaring the norm of both sides of Equation 74 of \cite{GK1}. Doing so, we find that $n_R^2 \tilde{P}_{\delta\phi} \propto P_{\delta\phi}\times (k_* a_R)^2 \propto P_{\delta\phi} \times J_R/(\kappa\lambda_*^2)$, so for the radial diffusion coefficient, we have
\begin{align}
    \label{eq:DRR_calibrated_global_long}
    D_{RR} \approx & \left(\frac{\overline{\Sigma}}{12\,\Msun/\mathrm{pc}^2}\right)^2 \left(\frac{\lambda_*}{600\,\mathrm{pc}}\right)^{n_\delta} \left(\frac{\tau_*}{70\,\mathrm{Myr}}\right) \nn \\
    & \times \left(\frac{\kappa}{40\,\mathrm{km/s/kpc}}\right)^{-1}\left(\frac{\xi}{0.94}\right)\left(\frac{\alpha_\mathrm{w}}{0.71}\right)^{-2} D_{RR}^{(R8)}.
\end{align}
Using the same Equation, we find $n_\varphi^2 \tilde{P}_{\delta\phi} \propto (k_* R)^2 P_{\delta\phi} \propto P_{\delta\phi} \times (\lambda_*/R)^2 $ since $n_\varphi = k_* R$, so for the azimuthal diffusion coefficient, we have
\begin{align}
    \label{eq:Dphiphi_calibrated_global_long}
    D_{\varphi\varphi} & \approx \left(\frac{\overline{\Sigma}}{12\,\Msun/\mathrm{pc}^2}\right)^2 \left(\frac{\lambda_*}{600\,\mathrm{pc}}\right)^{n_\delta} \left(\frac{\tau_*}{70\,\mathrm{Myr}}\right) \nn \\
    & \quad \times \left(\frac{R}{8\,\mathrm{kpc}}\right)^{2} \left(\frac{\xi}{0.91}\right)\left(\frac{\alpha_\mathrm{w}}{0.71}\right)^{-2} D_{\varphi\varphi}^{(R8)}.
\end{align}

In the short-wavelength regime $J_R \gg J_*$, we calculate $\tilde{P}_{\delta\phi}$ by squaring the norm of both sides of Equation B12 of \cite{GK1}, and using the large-argument Bessel-function asymptotic $J_{n_R}(k_* a_R) \propto (k_* a_R)^{-1/2}$. Consequently, we find that $n_R^2 \tilde{P}_{\delta\phi} \propto P_{\delta\phi}\times (k_* a_R)^{-1} \propto P_{\delta\phi} \times \lambda_* (\kappa/J_R)^{1/2}$, so for the radial diffusion coefficient, we have
\begin{align}
    \label{eq:DRR_calibrated_global_short}
    D_{RR} & \approx \left(\frac{\overline{\Sigma}}{12\,\Msun/\mathrm{pc}^2}\right)^2 \left(\frac{\lambda_*}{600\,\mathrm{pc}}\right)^{n_\delta + 3} \left(\frac{\tau_*}{70\,\mathrm{Myr}}\right) \nn \\
    & \quad \times \left(\frac{\kappa}{40\,\mathrm{km/s/kpc}}\right)^{\frac{1}{2}} \left(\frac{\xi}{0.68}\right)\left(\frac{\alpha_\mathrm{w}}{0.71}\right)^{-2} D_{RR}^{(R8)}.
\end{align}
Using the same Equation, we find $n_\varphi^2 \tilde{P}_{\delta\phi} \propto (k_* R)^2 P_{\delta\phi} \times (k_* a_R)^{-1} \propto P_{\delta\phi} \times \lambda_*^{-1} R^2 (\kappa/J_R)^{1/2}$, so for the azimuthal diffusion coefficient, we have
\begin{align}
    \label{eq:Dphiphi_calibrated_global_short}
    D_{\varphi\varphi} & \approx \left(\frac{\overline{\Sigma}}{12\,\Msun/\mathrm{pc}^2}\right)^2 \left(\frac{\lambda_*}{600\,\mathrm{pc}}\right)^{n_\delta + 1} \left(\frac{\tau_*}{70\,\mathrm{Myr}}\right) \nn \\
    & \quad \times \left(\frac{R}{8\,\mathrm{kpc}}\right)^{2} \left(\frac{\kappa}{40\,\mathrm{km/s/kpc}}\right)^{\frac{1}{2}} \nn \\
    & \quad \times \left(\frac{\xi}{0.82}\right)\left(\frac{\alpha_\mathrm{w}}{0.71}\right)^{-2} D_{\varphi\varphi}^{(R8)}.
\end{align}
We remind the reader that the explicit $J_R$-dependence of these calibrated diffusion coefficients is included in the measured expressions for $D_{ii}^{(R8)}$ (see \autoref{eq:DRR_R8} and \autoref{eq:Dphiphi_R8}).

Across all of these expressions, the suppression factor $\xi$ should be calculated following \autoref{eq:xi_definition}, and may vary with galactocentric radius through, e.g., $R$-dependent parameter values for the vertical gas density profile $\zeta$ (especially its scale height $H$), an $R$-dependent dominant wavelength $\lambda_*$, and/or an $R$-dependent vertical velocity dispersion $\sigma_{z}$. We also note that according to our idealized model, we expect the radial and azimuthal diffusion coefficients to share identical suppression factor values, but for self-consistency, have scaled \autoref{eq:DRR_calibrated_global_long} through \autoref{eq:Dphiphi_calibrated_global_short} by the values of $\xi$ measured in our fiducial simulations, which differ slightly (see \autoref{fig:vertical_suppression}). Neglecting possible small variations in $\xi$ between $D_{RR}$ and $D_{\varphi\varphi}$, we expect the constant heating-to-migration ratio achieved in the short-wavelength regime to satisfy (see \autoref{eq:heating_migration_ratio_short}):
\begin{equation}
    \label{eq:heating_migration_ratio_calibrated_global_short}
    \frac{\mathrm{rms} \,\delta J_R}{\mathrm{rms} \,\delta J_\varphi} \approx 0.055 \times \left(\frac{\lambda_*}{600\,\mathrm{pc}}\right) \left(\frac{R}{8\,\mathrm{kpc}}\right)^{-1}.
\end{equation}
Thus, we expect ISM-driven migration to remain ``cool,'' with $\mathrm{rms}\,\delta J_R\,/\,\mathrm{rms}\,\delta J_\varphi \ll 1$ over a wide range of galactic environments.

As a test of the scalings presented in \autoref{eq:DRR_calibrated_global_long} through \autoref{eq:Dphiphi_calibrated_global_short}, in \autoref{sec:LGR4} we report results from an alternative set of simulations using ISM gravitational potential fields from the TIGRESS-NCR LGR4 model, which uses a smaller nominal Galactocentric radius ($R_0 = 4\,$kpc) and a higher gas density ($\overline{\Sigma} = 39\,\Msun/\mathrm{pc}^2$) than the R8 simulations we have used so far. After measuring the diffusion coefficients from simulations with the same initial conditions considered in \autoref{sec:results} (see \autoref{eq:DRR_LGR4} and \autoref{eq:Dphiphi_LGR4}), we compare with the predicted values based on the scalings given above. We find reasonably good agreement, with relative errors at the level of a few tens of percent. More extensive environmental exploration of the amplitudes and scalings of the diffusion coefficients presented in \autoref{eq:DRR_calibrated_global_long} through
\autoref{eq:Dphiphi_calibrated_global_short} will be an important avenue for future work.

Finally, an alternative way to approach this problem is to use global simulations such as the NEXUS suite \citep{TepperGarcia2024, Zhang2025}, which include a prescription for modeling the dynamics of the gas (though in a way that is necessarily more coarse than TIGRESS-NCR) in conjunction with the stars. In such a simulation, one could in principle measure the power spectrum of gas fluctuations and link these measurements to the transport properties of the stars. Indeed, we are already encouraged by the fact that \cite{Zhang2025} find similar amounts of radial migration to us ($\approx 0.6\,$kpc for their $f_\mathrm{gas} = 20\%$ model after the $1.5$\,Gyr of evolution they simulate, see their Figure 2) and a heating-to-migration ratio comparable to our results ($\approx 0.08$). They also find that the amplitude of the migration scales approximately linearly with gas fraction, as anticipated by the scalings $D_{\varphi\varphi} \propto \overline{\Sigma}^2$ we predict (see their Figure 7). One could perform a series of systematic measurements to see how well these global results can be reproduced by ``gluing together'' the results from our local boxes at different galactocentric radii. We note that an interesting intermediate step between our local boxes and fully global approaches like those of \cite{Zhang2025} may be to study ISM conditions and stellar transport in TIGRESS boxes with embedded stellar spiral potentials \citep{KimKimOstriker2020}. These simulations retain the high spatial and temporal resolution and state-of-the-art prescriptions for gas physics applied in the R8 box of size $L_x = L_y = 1024\,$pc we study here, but probe a larger spatial domain of $L_x=\pi\,$kpc, $L_y = 2\pi\,$kpc, and would yield a better understanding of the interplay between ``ISM-driven'' and ``spiral-driven'' transport at larger scales, where the distinction between the two effects is unclear.

\subsection{Comparison with Milky Way Observations}
\label{sec:comparison_observations}

The goal of this paper is primarily to understand the physics underlying the radial heating and migration induced by a realistic ISM, rather than to model the ISM's full contribution to the AVR and migration history of the solar neighborhood. While in principle the latter endeavor may be approachable using the same local test-particle integration method described in \autoref{sec:rebound_simulations}, it would require a number of generalizations, such as (approximately in order from most to least significant):
\begin{enumerate}[label=(\roman*)]
    \item substantially time-varying ISM surface density to account for the more gas-rich Galactic environment at early times (as opposed to the nearly constant $\overline{\Sigma} = 12\,\Msun/\mathrm{pc}^2$ of the R8 model we assume here),
    \item realistic star formation histories (as opposed to the mono-age population we consider here), and
    \item time-dependent external potentials imposed by the growing stellar disk and dark matter halo (as opposed to the fixed potential specified by \autoref{eq:external_potential} we use here).
\end{enumerate}
Additionally, currently our stars are initialized completely agnostic to the underlying ISM structures, but the TIGRESS-NCR framework explicitly tracks clustered star formation in the form of star cluster particles that source stellar feedback, so stars could instead inherit their initial positions and velocities from their birth clusters. An interesting venue for future study is to assess the impact that these inhomogeneous initial conditions will have on ISM-driven dynamical perturbations, especially at early times when the ensemble is not yet fully phase-mixed.

Nevertheless, the results presented in \autoref{sec:results} enable us to place several informative constraints on the ISM's contribution to the transport observed in the solar neighborhood. In particular, the value of $\sigma_R \approx 17\,$km/s at  $t = 6\,$Gyr in our simulation is $\sim 40\%$ of the $\sigma_R \sim 40\,$km/s typically measured for stars with ages of $6\,$Gyr (see, e.g., \citealt{Mackereth2019} for an analysis using data from APOGEE DR14 and \textit{Gaia} DR2, and \citealt{Sharma2021} for an analysis using LAMOST DR4 and GALAH+DR3 data). The observed range of scalings $\sigma_R \propto t^{0.2-0.4}$ for stars with guiding center radii $R_g \in [7, 9]\,$kpc (see, e.g., Figure 6 of \citealt{Mackereth2019} or Figures 5 and 6 of \citealt{Sharma2021}) may plausibly be reproduced by a mixture of stars in the long- and short-wavelength regimes.\footnote{Because the transition from long- to short-wavelength regimes occurs at $\sigma_R \approx 6\,$km/s, we expect most stars to spend the vast majority of their lives in the short-wavelength regime.} The value of $\mathrm{rms}\,\delta x_\mathrm{g}\approx 0.73\,$kpc we measure at $t = 6\,$Gyr is also $\sim 30\%$ of the total amount of radial migration ($\mathrm{rms}\,\delta R_g \sim 2.5\,$kpc) measured for stars with ages of $\sim 6\,$Gyr in the solar neighborhood by, e.g., \cite{Frankel2020} using APOGEE DR14 and \textit{Gaia} DR2 data, or \cite{Zhang2025data} using LAMOST DR7 and \textit{Gaia} DR3 data. Additionally, the characteristic heating-to-migration ratio we measure, $\mathrm{rms}\,\delta J_R\,/\,\mathrm{rms}\,\delta J_\varphi \approx 0.055$, is somewhat lower than the value of $\sim 0.1$ measured by \cite{Frankel2020} and $\sim 0.08$ measured by \cite{Zhang2025data}.

Finally, we emphasize that because our simulations assume present-day solar-neighborhood conditions for the ISM, but the gas fraction was several times larger at early times (see (i) above), our measured total migration and velocity dispersion values may be viewed as \textit{lower bounds} on the ISM's possible contribution to the observed transport. Thus, the ISM contributes significantly to transport in the Galactic disk, and it is vital to adopt realistic models for its influence in future dynamical studies. Notably, from our experiments in varying the fluctuation amplitude (\autoref{fig:rmsdJ_t_variedA}), we see that while the amount of heating and migration scales linearly with fluctuation amplitude in the long-wavelength regime, the heating-to-migration ratio does not vary strongly with amplitude, so ISM-driven transport with a higher gas fraction could plausibly remain ``cool.'' It is reassuring that the heating-to-migration ratio we measure is smaller than suggested by observations, because our simulations neglect perturbers like spiral arms, the Galaxy's bar, and possible dark matter substructure, most of which are expected to drive that ratio higher. The lower bounds on ISM-driven transport we derive in this paper may therefore be used to place constraints on the amplitude, morphology, and dynamical history of all of these other perturbations to the Galactic disk.

\section{Summary}
\label{sec:summary}

In this paper, we have studied the radial heating and migration of stars driven by a realistic ISM using test-particle integrations in local shearing box models of a galactic disk. Our key findings are as follows:
\begin{itemize}
    \item ISM-driven radial transport in the solar neighborhood is predominantly driven by fluctuations of typical scale $\lambda_* \sim 600\,$pc (equivalently, $k_* \sim 0.01\,\mathrm{pc}^{-1}$), with a typical correlation time of $\tau_* \sim 70\,$Myr. One key quantity determining the radial transport behavior of an ensemble is $k_* a$, where $a$ is the typical (rms) epicyclic amplitude. Dynamically cold ensembles of stars reside in the long-wavelength regime $k_* a \ll 1$; all stars ultimately end up in the short-wavelength regime $k_* a \gg 1$ once their epicyclic amplitude is large enough. We expect the transition between these regimes to occur at $k_* a \sim 1$, which in our model means $a \sim 100\,$pc ($\sigma_R\sim 3$\,km/s), and indeed measure it to occur at $\sigma_R \approx 6\,$km/s in our simulations.
    \item For $k_* a \ll 1$ (long wavelengths, small epicycles, early times), transport proceeds as a random walk: radial velocity dispersion scales as $\sigma_R \propto t^{1/2}$, and radial migration scales as $\mathrm{rms}\,\delta J_\varphi \propto t^{1/2}$. The heating-to-migration ratio grows as $\mathrm{rms}\,\delta J_R\,/\,\mathrm{rms}\,\delta J_\varphi \propto t^{1/2}$.
    \item For $k_* a \gg 1$ (short wavelengths, larger epicycles, late times), stars cross multiple peaks and troughs of the ISM fluctuations as they traverse their epicycles, so that the kicks from each peak/trough partially cancel. In this regime, the transport is more gradual and can remain ``cool:'' radial velocity dispersion scales as $\sigma_R\propto t^{1/5}$, radial migration scales as $\mathrm{rms}\,\delta J_\varphi \propto t^{2/5}$, and the heating-to-migration ratio $\mathrm{rms}\,\delta J_R\,/\,\mathrm{rms}\,\delta J_\varphi$ is constant, typically $\ll 1$.
    \item This transport behavior can be modeled using the framework of quasilinear theory. \autoref{eq:DRR_R8} and \autoref{eq:Dphiphi_R8} give the action-space diffusion coefficients we measure in our fiducial solar-neighborhood simulations, which may be used in (semi-)analytic models of the ISM's diffusive dynamical effects (e.g., \autoref{eq:Langevin_update_step}). We do not yet have a precise understanding of how to embed our local, solar-neighborhood shearing box measurements in the global galactic context, but detail physically-motivated generalizations of our results in \autoref{eq:DRR_calibrated_global_long} through \autoref{eq:Dphiphi_calibrated_global_short}.
    \item We have confirmed that the ``cool'' transport effects described above can be reproduced using mock fluctuations generated using isotropic, \textit{spatio-temporal} Gaussian random field (GRF) models for the ISM, as predicted by quasilinear theory. Furthermore, we demonstrated that the temporal correlation structure of the ISM plays an important role in determining its transport behavior: ensembles experiencing mock GRF fluctuations generated with the correct \textit{spatial} power spectrum but white-noise temporal structure never transition to the short-wavelength regime scalings, leading to excess heating and erroneously ``hot'' migration.
    \item Only the amplitude, not the scalings, of the radial transport behavior depend on the vertical velocity dispersion of the stars; the reduction in transport that occurs in vertically warmer ensembles can be quantified with a simple suppression factor. So, ISM-driven transport is effectively separable into a two-dimensional radial problem and a one-dimensional vertical problem. We will study vertical ISM-driven transport in Paper II of this series (S. Modak et al., in preparation).
    \item In our fiducial solar-neighborhood simulations, after $10\,$Gyr, the final radial velocity dispersion is $\sigma_R \approx 18\,$km/s, the total radial migration is $\mathrm{rms}\,\delta x_\mathrm{g} \approx 0.88\,$kpc, and the heating-to-migration ratio is $\mathrm{rms}\,\delta J_R\,/\,\mathrm{rms}\,\delta J_\varphi \approx 0.055$. In our models with a higher fluctuation amplitude---closer to the conditions of the solar neighborhood in the past---the total amount of heating and migration is larger, but the heating-to-migration ratio is approximately the same. Therefore, the numbers quoted above are a \textit{lower bound} on the total contribution of the ISM to the observed radial transport in the Milky Way, and independent of its total contribution, ISM-driven migration remains ``cool.''
\end{itemize}
Overall, we emphasize that the dynamical effects of the ISM in our models are substantially different from those in conventional models of scattering driven by compact clouds. Specifically, realistic ISM fluctuations lead to distinct transport mechanisms, AVRs, and migration histories than classical theory would imply. These qualitative and quantitative differences suggest that conclusions from previous studies grounded in cloud-based models for the ISM, or those that neglect the ISM entirely, may need to be revised. The physical picture and semi-analytic prescriptions we develop here provide a useful starting point for future efforts to model and interpret the ISM's stellar-dynamical influence.

\begin{acknowledgments}
We thank Sanghyuk Moon, Chang-Goo Kim, Jenny Greene, and Jeremy Goodman for helpful conversations. S.M. acknowledges support from the National Science Foundation Graduate Research Fellowship under Grant No.\ DGE-2039656. C.H.\ is supported by the John N. Bahcall Fellowship Fund at the Institute for Advanced Study. Partial support for this work was provided by grant 510940 from the Simons Foundation to E.C.O.
\end{acknowledgments}

\appendix

\section{LGR4 Simulation Results}
\label{sec:LGR4}

In this Appendix, we present results of orbit integrations through ISM gravitational potential fields from the TIGRESS-NCR LGR4 model, instead of the R8 model used throughout the main text. The LGR4 simulation uses a nominal Galactocentric radius of $R_0 = 4\,$kpc, a box size of $L_x = L_y \equiv L = 512\,$pc and $L_z = 3072\,$pc, and a spatial resolution of $\Delta x = 2\,$pc. The circular orbital frequency is $\Omega = 30\,$km/s/kpc, the rotation curve is assumed to be flat ($q = 1$), and the externally imposed vertical potential due to stars and dark matter is again given by \autoref{eq:external_potential}, with parameters $\Sigma_* = 50\,\Msun/\mathrm{pc}^2$, $z_* = 500\,$pc, and $\rho_\mathrm{DM} = 0.005\,\Msun/\mathrm{pc}^3$. For reference, for these parameter choices, the vertical period is $2\pi/\nu \approx 113\,$Myr (see \autoref{eq:nu_TIGRESS}). We note that the LGR4 simulation is not meant to explicitly model a different region of the same galaxy as the R8 simulation. Instead, it serves as a model for ISM fluctuations in a more gas-rich environment: the simulation is initialized with a larger mean gas surface density of $\overline{\Sigma}(0) = 50\,\Msun/\mathrm{pc}^2$, which corresponds to a mean gas surface density of $\overline{\Sigma}\approx 39\,\Msun/\mathrm{pc}^2$ when a statistical steady-state is reached.\footnote{The substantial difference between the mean surface density in the LGR4 snapshots we analyze and the initial surface density (compared to the R8 simulation) is due to the $\sim 10$ times larger star formation rate in the LGR4 simulation (see Figure 1 of \citealt{Kim2023}).} In this work, as in \cite{Modak2026characterizing}, we use snapshots of the gravitational potential $\phi_\mathrm{g}$ taken every $1\,$Myr over the temporal baseline of $t \in [250\,\mathrm{Myr}, 350\,\mathrm{Myr}]$.

\begin{figure}
    \centering
    \includegraphics[width=0.47\textwidth]{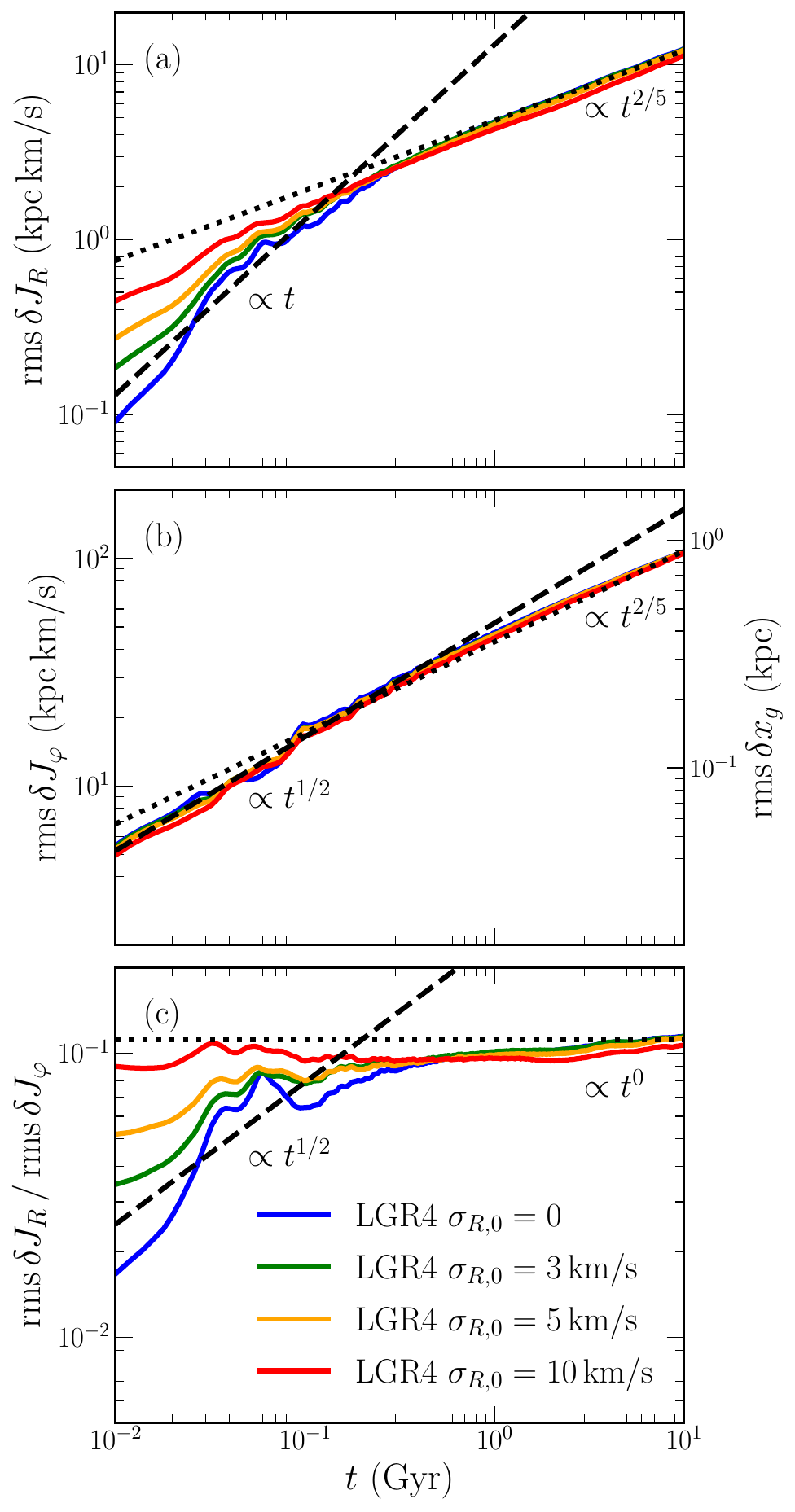}
    \caption{As in \autoref{fig:rmsdJ_t}, except for stellar ensembles integrated in ISM fluctuations drawn from the TIGRESS-NCR LGR4 simulation.}
    \label{fig:rmsdJ_t_LGR4}
\end{figure}

In \autoref{fig:rmsdJ_t_LGR4}, we show the results of simulations of $N = 50,000$ stellar orbits through the LGR4 gravitational potential fluctuations, carried out using the same methods described in \autoref{sec:rebound_simulations}, with $\sigma_{z,0} = 0$ for all ensembles, but varied $\sigma_{R, 0}$, analogous to the ensembles shown in \autoref{fig:rmsdJ_t}. Similar to the transport behavior observed in \autoref{fig:rmsdJ_t}, both $\mathrm{rms}\,\delta J_R$ and $\mathrm{rms}\,\delta J_\varphi$ grow more steeply than $\propto t^{2/5}$ before turning over to the more gradual growth rate, and correspondingly the heating-to-migration ratio converges to a near-constant value of $\mathrm{rms}\,\delta J_R\,/\,\mathrm{rms}\,\delta J_\varphi \approx 0.11$ at late times. The $\propto t^{2/5}$ scalings for radial heating and migration measured at late times are consistent with the predictions of the short-wavelength regime, and the $\propto t^{1/2}$ radial migration scaling measured at early times is also consistent with the long-wavelength regime as described in \autoref{sec:physical_interpretation}. The early-time radial heating scaling is less clear, and appears steeper than the $\propto t$ scaling predicted for the long-wavelength regime, but this may be attributed to the shorter duration of this regime compared to the R8 simulations---indeed, the steeper scalings here last for less than a single orbital time, and so the behavior may not be well-described by a diffusive approximation. Nevertheless, if we follow \autoref{sec:diffusion_measurements} in fitting power laws with indices predicted for each regime to the curves shown in panels (a) and (b) of the Figure, using \autoref{eq:rms_dJR_approx} and \autoref{eq:rms_dJphi_approx}, we measure
\begin{equation}
    \label{eq:DRR_LGR4}
    D_{RR}^{\mathrm{(LGR4)}} \! \approx \! \begin{cases}
        18\,\frac{\mathrm{(kpc\,km/s)^2}}{\mathrm{Gyr}}\times\left(\frac{J_R}{\mathrm{kpc\,km/s}}\right) & \! \! (J_R \ll J_{*}) \\[4pt]
        9.6\,\frac{\mathrm{(kpc\,km/s)^2}}{\mathrm{Gyr}}\times\left(\frac{J_R}{\mathrm{kpc\,km/s}}\right)^{-\frac{1}{2}} & \! \! (J_R \gg J_{*})
    \end{cases}
\end{equation}
for the radial diffusion coefficient, and
\begin{equation}
    \label{eq:Dphiphi_LGR4}
    D_{\varphi\varphi}^{\mathrm{(LGR4)}} \! \approx \! \begin{cases}
        2700\,\frac{\mathrm{(kpc\,km/s)^2}}{\mathrm{Gyr}} & (J_R \ll J_{*}) \\[4pt]
        1500\,\frac{\mathrm{(kpc\,km/s)^2}}{\mathrm{Gyr}} \! \times \! \left( \! \frac{J_R}{\mathrm{kpc\,km/s}} \! \right)^{-\frac{1}{2}} & (J_R \gg J_{*})
    \end{cases}
\end{equation}
for the azimuthal diffusion coefficient.

To compare these measured values with the our fiducial results for the R8 simulation, we now require a quantitative characterization of the ISM fluctuations; we refer the interested reader to \cite{Modak2026characterizing}, and just summarize the features relevant to radial transport here. The spatio-temporal spectrum of linear surface density fluctuations takes a similar shape to that of the R8 simulation: the spatial component is well-described by a power law with an index of $n_\delta = 2.2$, and the temporal spectrum is flat at low frequencies before declining sharply at $\omega_c(k) = 0.4\omega_0(k)$, where $\omega_0(k) = [\tau_0^{-2} + v_\mathrm{eff}^2k^2]^{1/2}$ with $\tau_0 = 5\,$Myr just like in the R8 simulation, but with $v_{\mathrm{eff}} = 10\,$km/s (instead of $12\,$km/s for R8). Unlike the R8 simulation, the spatial spectrum does not show any indications of a turnover, so the dominant fluctuations are set by the box size to be $k_* = 2\pi/L \approx 0.012\,\mathrm{pc}^{-1}$ (or $\lambda_* = 512\,$pc). Following \autoref{eq:taustar_definition}, the corresponding correlation time is still $\tau_* \sim 70\,$Myr. The vertical structure of the gas layer is well-described by the same dimensionless profile $\zeta$ given in \autoref{eq:zeta_definition}, but with $\alpha_\mathrm{s} = 1.02$, $\alpha_\mathrm{e} = 2.6$, and $w = 0.07$, which corresponds to $\alpha_\mathrm{w} = 1.1$. The mean rms thickness of the gas layer is $H = 165\,$pc, but varies over time between $\approx 100-200\,$pc across the snapshots we use.

Using the values of $\overline{\Sigma}$, $\lambda_*$, $\tau_*$, $R$, $\alpha_\mathrm{w}$, and $H$ described in the preceding paragraphs for the LGR4 fluctuations, \autoref{eq:DRR_calibrated_global_long} and \autoref{eq:DRR_calibrated_global_short} predict
\begin{equation}
    \label{eq:DRR_LGR4_predicted}
    D_{RR}^{\text{(pred.)}} \! \approx \! \begin{cases}
        21\,\frac{\mathrm{(kpc\,km/s)^2}}{\mathrm{Gyr}}\times\left(\frac{J_R}{\mathrm{kpc\,km/s}}\right) & \! \! (J_R \ll J_{*}) \\[4pt]
        7.7\,\frac{\mathrm{(kpc\,km/s)^2}}{\mathrm{Gyr}}\times\left(\frac{J_R}{\mathrm{kpc\,km/s}}\right)^{-\frac{1}{2}} & \! \! (J_R \gg J_{*})
    \end{cases}
\end{equation}
for the radial diffusion coefficient, while \autoref{eq:Dphiphi_calibrated_global_long} and \autoref{eq:Dphiphi_calibrated_global_short} predict
\begin{equation}
    \label{eq:Dphiphi_LGR4_predicted}
    D_{\varphi\varphi}^{\text{(pred.)}} \! \approx \! \begin{cases}
        4400\,\frac{\mathrm{(kpc\,km/s)^2}}{\mathrm{Gyr}} & \! \! (J_R \ll J_{*}) \\[4pt]
        1500\,\frac{\mathrm{(kpc\,km/s)^2}}{\mathrm{Gyr}}\times\left(\frac{J_R}{\mathrm{kpc\,km/s}}\right)^{-\frac{1}{2}} & \! \! (J_R \gg J_{*})
    \end{cases}
\end{equation}
for the azimuthal diffusion coefficient. In both these estimates, we have made the simplifying assumption that $n_\delta = 2.3$ for both the R8 and LGR4 simulations (which is reasonable, given that both quantities are reported with error bars of $\pm 0.1$ in Table 2 of \citealt{Modak2026characterizing}). We have also evaluated the vertical suppression factor $\xi$ assuming a constant $H = 165\,$pc, using time-averaged values of $\sigma_z \approx 3.9\,$km/s in the long-wavelength regime and $\sigma_z \approx 11\,$km/s in the short-wavelength regime, which we measure directly from our ensembles. We see that the measured diffusion coefficient values given in \autoref{eq:DRR_LGR4} and \autoref{eq:Dphiphi_LGR4} are reasonably consistent with our predictions using the mean value of $H$, with relative errors of on average $25\%$, and note that they fall well within the range of values we would predict had we instead set $H = 100\,$pc or $H = 200\,$pc in calculating the vertical suppression.

Following \autoref{eq:heating_migration_ratio_calibrated_global_short}, the predicted heating-to-migration ratio in the short-wavelength regime is $\mathrm{rms}\,\delta J_R\,/\,\mathrm{rms}\,\delta J_\varphi \approx 0.094$, within $15\%$ of the measured value of $0.11$. Finally, using \autoref{eq:Jstar_calibrated}, the characteristic radial action value separating the asymptotic long- and short-wavelength regimes is $J_* \approx 0.16\,\mathrm{kpc}\,$km/s. In a broken power-law model for the diffusion coefficients measured in \autoref{eq:DRR_LGR4} and \autoref{eq:Dphiphi_LGR4} above, the transition between regimes occurs at $J_R \approx 0.65\,\mathrm{kpc}\,$km/s for $D_{RR}^{\mathrm{(LGR4)}}$, and at $J_R \approx 0.32\,\mathrm{kpc}\,$km/s for $D_{\varphi\varphi}^{\mathrm{(LGR4)}}$. Note that the transition value is $\sim 4J_*$ for the radial diffusion coefficient and $\sim 2J_*$ for the azimuthal diffusion coefficient, just as found for the R8 simulations in \autoref{sec:diffusion_measurements}.

\section{REBOUND Integrator Details}
\label{sec:REBOUND_details}

In this Appendix, we describe the integration routine we use to solve the equations of motion (\autoref{eq:shearing_eom_x} through \autoref{eq:shearing_eom_z}) in greater detail. Our method is based on the symplectic epicycle integrator (SEI) derived and implemented in \texttt{REBOUND} as described in \cite{ReinTremaine2011}, and we follow their notation below. To evolve a particle from time $t^{(n)}$ to time $t^{(n+1)} \equiv t^{(n)} + \Delta t$, writing $Q^{(n)}$ or $Q^{(n+1)}$ correspondingly for the value of a given quantity $Q$ at each time, we use a mixed-variable, symplectic ``drift-kick-drift'' (DKD) scheme as follows:
\begin{enumerate}[label=(\roman*)]
    \item $\hat{H}_0 \left(\frac{\Delta t}{2}\right) \! : \!  (\bx^{(n)}, \bvee^{(n)}) \mapsto (\bx^{(n+\frac{1}{2}, +)}, \bvee^{(n+\frac{1}{2}, -)})$,
    \item $\hat{\Phi}(\Delta t) \! : \!  \bvee^{(n+\frac{1}{2}, -)} \mapsto \bvee^{(n+\frac{1}{2}, +)}$, and
    \item $\hat{H}_0 \left(\frac{\Delta t}{2}\right) \! : \!  (\bx^{(n+\frac{1}{2}, +)}, \bvee^{(n+\frac{1}{2}, +)}) \mapsto (\bx^{(n+1)}, \bvee^{(n+1)})$.
\end{enumerate}
In this procedure, the drift operator $\hat{H}_0(\Delta t)$ accounts for the unperturbed epicyclic planar motion as specified in \autoref{eq:shearing_eom_unperturbed_x} and \autoref{eq:shearing_eom_unperturbed_y}, as well as the vertical motion in the TIGRESS-NCR vertical potential given in \autoref{eq:external_potential}. The kick operator $\hat{\Phi}(\Delta t)$ accounts for the external forcing by the ISM; post-kick quantities are marked with a $+$ in the upper index while pre-kick quantities are marked with a $-$. We describe the drift operator in \autoref{sec:REBOUND_details_drift}, and the kick operator in \autoref{sec:REBOUND_details_kick}.

\subsection{Drift Operator}
\label{sec:REBOUND_details_drift}

The update rules resulting from applying $\hat{H}_0(\Delta t)$ on a particle at phase space position $(\bx^{(n)}, \bvee^{(n)})$ to evolve it to a new position $(\bx^{(n+1)}, \bvee^{(n+1)})$ are described in Equation 9 through Equation 13 of \cite{ReinTremaine2011} for the Keplerian context, where $q = 3/2$ and $\kappa = \Omega$, and motion in the vertical plane is also assumed to be exactly epicyclic with $\nu = \Omega$. Below, we write generalizations of these update rules for arbitrary $0 < q < 2$ (with $\kappa \equiv \sqrt{2(2-q)}\Omega$), and for an arbitrary vertical potential.

For the planar variables, the generalization of Equation 9 of \cite{ReinTremaine2011} is
\begin{align}
    \label{eq:H0_operator_start}
    x_0^{(n)} & = \frac{1}{2-q}\left(\frac{v_y^{(n)}}{\Omega} + 2x^{(n)}\right), \nn \\
    y_0^{(n)} & = y^{(n)} - \frac{1}{2-q}\frac{v_x^{(n)}}{\Omega},
\end{align}
the generalization of their Equation 10 is
\begin{align}
    x_s^{(n)} & = \Omega (x^{(n)} - x_0^{(n)}), \nn \\
    y_s^{(n)} & = \frac{\kappa}{2}(y^{(n)} - y_0^{(n)}),
\end{align}
the generalization of their Equation 11 is\footnote{We also follow Appendix A of \cite{ReinTremaine2011} in the implementation of the rotation step in \autoref{eq:rotation_step} (their Equation 11) as three shear operators for improved precision.}
\begin{align}
    \label{eq:rotation_step}
    x_s^{(n+1)} & = x_s^{(n)}\cos(\kappa\Delta t) + y_s^{(n)}\sin(\kappa\Delta t), \nn \\
    y_s^{(n+1)} & = -x_s^{(n)}\sin(\kappa \Delta t) + y_s^{(n)}\cos(\kappa\Delta t),
\end{align}
and the generalization of their Equation 12 is
\begin{align}
    x^{(n+1)} & = \frac{x_s^{(n+1)}}{\Omega} + x_0^{(n)}, \nn \\
    y^{(n+1)} & = \sqrt{\frac{2}{2-q}}\frac{y_s^{(n+1)}}{\Omega} + y_0^{(n)} - q\Omega x_0^{(n)}\Delta t, \nn \\
    v_x^{(n+1)} & = \sqrt{2(2-q)}y_s^{(n+1)}, \nn \\
    v_y^{(n+1)} & = -2x_s^{(n+1)} - q\Omega x_0^{(n)}.
\end{align}
We have checked that following these update rules reproduces analytic expectations from \autoref{eq:shearing_eom_unperturbed_x} and \autoref{eq:shearing_eom_unperturbed_y} up to to machine precision. Note that \cite{Quillen2018} arrived at similar update rules for modifying the SEI to galactic contexts, but their Equation A6, Equation A7, and Equation A9 contain typos and do not produce the correct orbits as written.

For the vertical variables, because orbits in the potential \autoref{eq:external_potential} do not have a closed-form solution, in place of Equation 13 of \cite{ReinTremaine2011} which assumes vertical epicyclic motion, we use an additional DKD sub-step, which proceeds as follows:
\begin{align}
    \label{eq:H0_operator_end}
    z_d & = z^{(n)} + v_z^{(n)}\frac{\Delta t}{2}, \nn \\
    v_z^{(n+1)} & = v_z^{(n)} - \frac{\md \phi_\mathrm{vert}}{\md z}\bigg|_{z = z_d}\Delta t, \nn \\
    z^{(n+1)} & = z_d + v_z^{(n+1)}\frac{\Delta t}{2}.
\end{align}
We have checked that following these update rules leads to bounded (oscillatory) energy errors at late times as expected for a symplectic integrator, and that the resulting orbits are indeed nearly harmonic with frequency given by \autoref{eq:nu_TIGRESS} if the vertical excursions from the midplane remain small.

Finally, we remind the reader that the SEI involves the operator $\hat{H}_0(\Delta t/2)$, not $\hat{H}_0(\Delta t)$. Above, we have written the update rules for a full timestep $\Delta t$ following the convention of \cite{ReinTremaine2011}, for clarity and to allow for a more direct comparison with their method. When implementing the update rules numerically in steps (i) and (iii) of the DKD scheme, $\Delta t$ should be replaced by $\Delta t/2$ everywhere in \autoref{eq:H0_operator_start} through \autoref{eq:H0_operator_end}. Correspondingly, throughout these equations, the upper indices should be written as $(n)$ and $(n+1/2, \pm)$ for step (i), and $(n+1/2, +)$ and $(n+1)$ for step (iii).

\subsection{Kick Operator}
\label{sec:REBOUND_details_kick}

Our implementation of the update rule for the kick operator $\hat{\Phi}(\Delta t)$ is identical to that described in Equation 15 of \cite{ReinTremaine2011}, but we repeat it here in our notation for completeness:
\begin{equation}
    \label{eq:kick_operator}
    \bvee^{(n+1, +)} = \bvee^{(n+\frac{1}{2}, -)} - \nabla \phi_\mathrm{g}(\bx^{(n+\frac{1}{2}, +)}, t^{(n+\frac{1}{2})}) \Delta t,
\end{equation}
i.e., $\hat{\Phi}(\Delta t)$ acts as a pure velocity kick, where $\nabla\phi_\mathrm{g}(\bx^{(n+\frac{1}{2}, +)}, t^{(n+\frac{1}{2})})$ is the acceleration induced by the ISM gravitational potential evaluated at the intermediate position of the particle after one application of the drift operator $\hat{H}_0(\Delta t/2)$, and at time $t^{(n+\frac{1}{2})} \equiv t^{(n)} + \Delta t/2$. We implement this operation as an ``external force'' in \texttt{REBOUND} as follows.

We begin the simulation by loading into memory the first two TIGRESS-NCR snapshots, which correspond to the times $t^{(i)} = 250\,$Myr and $t^{(f)} \equiv t^{(i)} + 1\,\mathrm{Myr} = 251\,$Myr, and label these snapshots $\phi_\mathrm{g}^{(i)}$ and $\phi_\mathrm{g}^{(f)}$ respectively. At each time $t^{(n+\frac{1}{2})}$ at which we want to access the potential, we first check whether $t^{(n+\frac{1}{2})} > t^{(f)}$: in that case, we update the value of $t^{(i)}$ to $t^{(f)}$, increment the value of $t^{(f)}$ by 1\,Myr, replace the snapshot $\phi_\mathrm{g}^{(i)}$ with the previous $\phi_\mathrm{g}^{(f)}$, and load in the next snapshot as the new $\phi_\mathrm{g}^{(f)}$. Thus, at each time, we store only two TIGRESS-NCR snapshots in memory, and only load in new snapshots on the fly as needed. Once we have snapshots and time labels such that $t^{(i)} < t^{(n+\frac{1}{2})} \leq t^{(f)}$, we linearly interpolate the potential between the snapshots according to
\begin{align}
    \label{eq:interpolated_phi}
    \phi_\mathrm{g}(\bx, t^{(n + \frac{1}{2})}) & \equiv \left(\frac{t^{(n+\frac{1}{2})} - t^{(i)}}{1\,\mathrm{Myr}}\right)\phi_\mathrm{g}^{(f)}(\bx) \nn \\
    & \quad\quad\quad\quad + \left(\frac{t^{(f)} - t^{(n+\frac{1}{2})}}{1\,\mathrm{Myr}}\right)\phi_\mathrm{g}^{(i)}(\bx).
\end{align}
Next, for each particle, we identify the nearest spatial cell center to its position $\bx^{(n+\frac{1}{2}, +)}$, which we denote $\overline{\bx}$, and approximate the gradient $\nabla \phi_\mathrm{g}(\bx^{(n+\frac{1}{2}, +)}, t^{(n+\frac{1}{2})})$ using centered second-order differences at $\overline{\bx}$. For instance,
\begin{equation}
    \label{eq:second_order_difference_centered}
    \frac{\partial\phi_\mathrm{g}}{\partial y}(\bx^{(n+\frac{1}{2}, +)}, t^{(n + \frac{1}{2})}) \approx \frac{\phi_\mathrm{g}(\overline{y} + \Delta x) - \phi_\mathrm{g}(\overline{y} - \Delta x)}{2\Delta x},
\end{equation}
where we have suppressed the $(x = \overline{x}, z=\overline{z}, t = t^{(n + \frac{1}{2})})$ arguments in $\phi_\mathrm{g}$ for brevity. At the box boundaries in $y$, we can make use of the periodic boundary conditions to continue to apply \autoref{eq:second_order_difference_centered}, adding or subtracting $L$ as needed so that $\overline{y}\pm\Delta x$ falls inside the box. At the box boundaries in $x$, we could similarly continue to apply \autoref{eq:second_order_difference_centered} by making use of the \textit{shearing}-periodic boundary conditions (see, e.g., Equation 1 through Equation 3 of \citealt{Modak2026characterizing}), but the additional coordinate transformations required increase the computational cost, so we instead apply forward (upper signs) and backward (lower signs) differences: for particles whose nearest cell center is at $\overline{x}=(\mp L \pm \Delta x)/2$, we approximate the gradients as
\begin{align}
    \label{eq:second_order_difference_forwardbackward}
    \frac{\partial\phi_\mathrm{g}}{\partial x}(\bx^{(n+\frac{1}{2}, +)}, t^{(n + \frac{1}{2})}) \! \approx \! \frac{1}{2\Delta x}\bigg[ & \mp 3\phi_\mathrm{g}(\overline{x}) \pm 4\phi_\mathrm{g}(\overline{x} \pm \Delta x) \nn \\
    & \quad\quad \mp \phi_\mathrm{g}(\overline{x} \pm 2\Delta x)\bigg],
\end{align}
where we have again suppressed the other arguments of $\phi_\mathrm{g}$ for brevity. The box boundaries in $z$ are essentially never reached by any particles, but we implement analogues of the forward/backward differences in \autoref{eq:second_order_difference_forwardbackward} when $\overline{z}=(\mp L_z \pm \Delta x)/2$ as well.

Finally, because the temporal baseline of the available TIGRESS-NCR snapshots is much less than the $10\,$Gyr duration over which we would like to study the dynamical effects of the ISM, we loop the TIGRESS-NCR snapshots repeatedly over the course of the simulation. Concretely, once $t^{(f)}$ reaches the end of the temporal baseline of TIGRESS-NCR snapshots ($t^{(f)} = 450\,$Myr for the R8 simulation or 350\,Myr for the LGR4 simulation), at the first instance when $t^{(n + \frac{1}{2})} > t^{(f)}$, we reset the values $t^{(i)} = 250\,$Myr and $t^{(f)} = 251\,$Myr, and replace $\phi_\mathrm{g}^{(i)}$ and $\phi_\mathrm{g}^{(f)}$ with the corresponding snapshots as well. This way, we avoid spurious interpolation between the final and initial TIGRESS-NCR snapshots in the temporal baseline; the single instantaneous discontinuity at the time the loop resets has no major dynamical effects. To verify this explicitly, we have checked that using a shorter loop period of 100\,Myr instead of 200\,Myr in orbit integrations using the R8 simulation snapshots (i.e., resetting $t^{(i)}$, $t^{(f)}$, $\phi_\mathrm{g}^{(i)}$, and $\phi_\mathrm{g}^{(f)}$ after $t^{(f)} = 350\,$Myr instead of 450\,Myr) does not meaningfully alter our results. We have also found good agreement between our fiducial simulations and orbit integrations using fluctuations drawn from a lower resolution TIGRESS-NCR R8 model (with $\Delta x = 8\,\mbox{pc}$) for which snapshots spanning a baseline of $t \in [250\,\mathrm{Myr}, 650\,\mathrm{Myr}]$ are available, enabling a test with a longer loop period of 400\,Myr. 

Of course, this loop of simulation snapshots that are representative of the solar-neighborhood ISM \textit{today} lacks many aspects of a realistic ISM over the secular timescales of interest, such as a larger initial gas surface density that approaches its present value over several Gyr of evolution. However, as described in \autoref{sec:comparison_observations}, here we focus on building a physical understanding of the transport driven by a realistic ISM, and for that purpose, the use of this loop is well-motivated. The ISM in the TIGRESS-NCR simulation is in statistical steady state over the temporal baseline we use, and the loop times of $200\,$Myr and $100\,$Myr for the R8 and LGR4 simulations respectively are longer than the correlation time $\tau_*$ for the dominant set of fluctuations in each simulation (see \autoref{eq:taustar_definition}). We defer a study of more realistic ISM-driven transport using fluctuations that evolve over secular timescales (in place of a na\"ive looping procedure as carried out here) to future work.

\section{Relating rms action changes to diffusion coefficients}
\label{sec:rms_from_D}

In this Appendix, we relate the diffusion coefficients $D_{RR}$ and $D_{\varphi\varphi}$ to the rms action changes, under certain assumptions which we will outline below as required.

We start with radial heating: the mean radial action is given by
\begin{equation}
    \label{eq:meanJR}
    \langle J_R \rangle \equiv \frac{(2\pi)^2}{M} \int \md \bJ\,J_R\,f_0(\bJ, t),
\end{equation}
where $M$ is the mass in stars enclosed in the action-space region of interest. Taking the time derivative and substituting the kinetic equation given in \autoref{eq:quasilinear_diffusion} into the right hand side, integrating by parts, and ignoring any boundary terms, we soon arrive at 
\begin{align}
    \label{eq:heating_rate}
    \frac{\md \langle J_R \rangle}{ \md t} 
    &= \frac{(2\pi)^2}{M} \int \md \bJ  \frac{1}{2}\left[ \frac{\partial D_{\varphi R }}{\partial J_\varphi} + \frac{\partial D_{RR }}{\partial J_R} \right] f_0.
\end{align}
For the homogeneous fluctuations in a local shearing box we consider, there is no special value of $J_\varphi$, so the diffusion coefficients $D_{ij}$ must be independent of $J_\varphi$, and the first term in the large bracket is identically zero. Indeed, the first term is typically smaller than the second term by $\mathcal{O}(\langle J_R\rangle/J_\varphi)$, so it is often ignorable even in broader contexts---for an explicit demonstration of this fact, see C. Hamilton et al. (in preparation).

If we further assume that $D_{RR}(\bJ)$ has a power-law dependence on $J_R$ as in \autoref{eq:DRR_power}, \textit{and} that the DF retains the standard Schwarzschild exponential dependence on $J_R$ of \autoref{eq:schwarzschild_IC} except possibly with a time-dependent mean $\langle J_R \rangle$, we can perform the action-space integral to find
\begin{equation}
      \frac{\md \langle J_R \rangle}{ \md t} \simeq \frac{1}{2}\Gamma(1 + \alpha_R) d_{RR} \langle J_R  \rangle^{\alpha_R-1}.
      \label{eq:heating_rate_power}
\end{equation}
For $\alpha_R\neq 2$ (which always holds in the physically relevant cases), this equation has the solution 
\begin{equation}
    \label{eq:mean_JR_evolution}
    \langle J_R \rangle = \big[ \langle J_R \rangle_0^{2-\alpha_R} + \left(1-\frac{\alpha_R}{2}\right)\Gamma(1 + \alpha_R)\,d_{RR} \, t \big]^{1/(2-\alpha_R)}.
\end{equation}
If the initial DF is sufficiently cold, $\langle J_R\rangle_0$ will be negligible, and we can write $\mathrm{rms}\,\delta J_R \simeq \langle J_R^2 \rangle^{1/2} \simeq \sqrt{2}\langle J_R\rangle$, where the last approximate equality follows from the Schwarzchild DF assumption. With this, defining
\begin{equation}
    \label{eq:c_alphaR_definition}
    c(\alpha_R) \equiv \left[\left(1 - \frac{\alpha_R}{2}\right)\Gamma(1+\alpha_R)\right]^{1/(2-\alpha_R)},
\end{equation}
we arrive at \autoref{eq:rms_dJR_approx}.

To study radial migration, we consider not the full action-space DF $f_0(\bJ, t)$ but rather the marginalized angular momentum DF $F_0(J_\varphi, t) = 2\pi \int \md J_R \, f_0$ (see Equation 24 of \citealt{GK1}). Integrating \autoref{eq:quasilinear_diffusion} with respect to $J_R$, we find that this marginalized DF satisfies
\begin{equation}
    \label{eq:dFdt_general}
    \frac{\p F_0}{\p t} = \pi \frac{\p}{\p J_\varphi} \int_0^\infty \md J_R\,\bigg(
    D_{\varphi\varphi}\frac{\p f_0}{\p J_\varphi } + D_{\varphi R}\frac{\p f_0}{\p J_R}
    \bigg).
\end{equation}
It is often (but certainly not always) the case that the second term on the right hand side of \autoref{eq:dFdt_general} is subdominant. Briefly, inspecting \autoref{eq:D_from_power}, the (typically large) $n_R = 0$ component of the fluctuation spectrum $\tilde{P}_{\delta\phi}$ is absent from $D_{\varphi R}$, because it is multiplied by the prefactor $n_R n_\varphi = 0$ in the sum, but it is present in $D_{\varphi\varphi}$ because the prefactor is instead $n_\varphi^2$. Furthermore, components of the fluctuation spectrum with $n_R \geq 1$ do not contribute to the diffusion because there is limited fluctuation power at the corresponding frequencies $\omega_\mathrm{res} = n_R\kappa + n_\varphi(\Omega_\varphi - \Omega)$. Again, for a more detailed explanation, see C. Hamilton et al. (in preparation).

If we neglect the second term for simplicity, and if we assume that $D_{\varphi\varphi}$ has power-law dependence on $J_R$ as in \autoref{eq:Dphiphi_power}, \textit{and} that $f_0$ is of Schwarzchild form, then \autoref{eq:dFdt_general} reduces to
\begin{equation}
    \label{eq:dFdt_diffusion}
    \frac{\p F_0}{\p t} \simeq  \frac{1}{2}\frac{\p}{\p J_\varphi}   \langle D_{\varphi\varphi} \rangle \frac{\p F_0}{\p J_\varphi},
\end{equation}
where 
\begin{align}
    \label{eq:D_0_definition}
    \langle D_{\varphi\varphi} \rangle &\equiv  \Gamma(1+ \alpha_\varphi) d_{\varphi\varphi} \langle J_R\rangle^{\alpha_\varphi} \nn \\
    & = \Gamma(1+\alpha_\varphi)d_{\varphi\varphi}c(\alpha_R)^{\alpha_\varphi} (d_{RR}t)^{\alpha_\varphi/(2-\alpha_R)}
\end{align}
is just the phase space average of $D_{\varphi\varphi}(\bJ)$ taken over a narrow $J_\varphi$ range, and to get the second line we substituted the radial heating result of \autoref{eq:mean_JR_evolution} (and again used the assumption that the initial value of $\langle J_R \rangle$ is negligible).

Since there are no special angular momenta in our problem, \autoref{eq:dFdt_diffusion} is a diffusion equation whose diffusion coefficient depends only on time. It can therefore be solved analytically. In general, the equation 
\begin{equation}
    \frac{\partial F(J,t)}{\partial t} = \frac{1}{2}\frac{\partial}{\partial J} (Ct^n) \frac{\partial F}{\partial J}
\end{equation}
for constant $C$ has the solution
\begin{equation}
    \label{eq:F0_solution_pure_diffusion}
    F(J, t) = \int \md J' \, F(J', 0) \,\mathcal{G}(J - J', t),
\end{equation}
where the Green's function $\mathcal{G}$ is a Gaussian,
\begin{equation}
   \mathcal{G}(J, t) = \frac{1}{\sqrt{2\pi\mathcal{S}^2(t)}} \me^{-J^2/(2\mathcal{S}^2(t))},
    \label{eq:green_function_pure_diffusion}
\end{equation}
with time-dependent variance\footnote{An important special case is when $n=0$. Then the diffusion coefficient has no time dependence, and the Green's function reduces to the standard spreading Gaussian with variance $2Ct$.}
\begin{equation}
    \mathcal{S}^2(t) \equiv \frac{2C}{n + 1}t^{n + 1}.
\end{equation}
Over a narrow range of $J$, then, the root mean square change is approximately given by rms\,$\delta J(t) = \mathcal{S}(t)$. By comparing this solution with the diffusion coefficient given in \autoref{eq:D_0_definition}, we can read off the rms change in angular momentum to be \autoref{eq:rms_dJphi_approx} as written in the main text.

\bibliography{bibliography}{}
\bibliographystyle{aasjournalv7}

\end{document}